\documentclass[fleqn,usenatbib]{mnras}

\usepackage{xcolor}

\usepackage{longtable}
\usepackage{deluxetable} 
\RequirePackage{rotating} 

\usepackage[T1]{fontenc}

\DeclareRobustCommand{\VAN}[3]{#2}
\let\VANthebibliography\thebibliography
\def\thebibliography{\DeclareRobustCommand{\VAN}[3]{##3}\VANthebibliography}

\usepackage{graphicx}   
\usepackage{amsmath}    
\usepackage{amssymb}    

\newcommand{\RN}[1]{\MakeUppercase{\romannumeral #1}}

\title[Candidate Dusty Starbursts at $z>6$]{Counterparts of Candidate Dusty Starbursts at $z>6$}

\author[H. Yan et al.]{
Haojing Yan,$^{1}$\thanks{E-mail: yanha@missouri.edu}
Chenxiaoji Ling,$^{1}$
and Zhiyuan Ma$^{2}$
\\
$^{1}$Department of Physics and Astronomy, University of Missouri-Columbia, Columbia, MO 65203, USA\\
$^{2}$Department of Astronomy, University of Massachusetts, Amherst, MA 01003, USA 
}

\date{Accepted XXX. Received YYY; in original form ZZZ}

\pubyear{2022}

\begin{document}
\label{firstpage}
\pagerange{\pageref{firstpage}--\pageref{lastpage}}
\maketitle

\begin{abstract}
  
    We present an analysis of the optical-to-near-IR counterparts of a sample 
of candidate dusty starbursts at $z>6$. These objects were pre-selected based
on the rising trend of their far-infrared-to-sub-millimeter spectral energy 
distributions and the fact that they are radio-weak. Their precise positions
are available through millimeter and/or radio interferometry, which enable us
to search for their counterparts in the deep optical-to-near-IR images. The
sample include five $z>6$ candidates. Three of them have their counterparts 
identified, one is still invisible in the deepest images, and one is a known
galaxy at $z=5.667$ that is completely blocked by a foreground galaxy. The 
three with counterparts identified are analyzed using population systhesis
model, and they have photometric redshift solutions ranging from 7.5 to 9.0.
Assuming that they are indeed at these redshifts and that they are not 
gravitationally lensed, their total IR luminosities are
$10^{13.8-14.1} L_\odot$ and the inferred star formation rates are
6.3--13~$\times 10^3$~$M_\odot$~yr$^{-1}$. The existence of dusty starbursts at
such redshifts would imply that the universe must be forming stars intensely
very early in time in at least some galaxies, otherwise there would not be
enough dust to produce the descendants observed at these redshifts.
The inferred host galaxy stellar masses of these three objects, which are at
$\gtrsim 10^{11} M_\odot$ (if not affected by gravitational lensing), present
a difficulty in explanation unless we are willing to accept that their 
progenitors either kept forming stars at a rate of 
$\gtrsim 10^3$~$M_\odot$~yr$^{-1}$ or were formed through intense instantaneous
bursts. Spectroscopic confirmation of such objects will be imperative.

\end{abstract}


\section{Introduction}

    Ultra-luminous Infrared Galaxies (ULIRGs) have extremely high infrared 
luminosity of $L_{\rm IR}\geq 10^{12} L_\odot$ (integrated over the rest-frame
8 to 1000 $\mu$m). It is widely believed that they are starbursts enshrouded by
dust, with star formation rates (SFRs) of $\geq 100 M_\odot$~yr$^{-1}$. The 
intense UV radiation from young stars in the starburst region is absorbed by
dust, which is re-radiated in infrared (IR) and gives rise to the 
enormous IR luminosity.

    ULIRGs must be heavily metal-enriched because metal is needed to form dust. 
It is therefore somewhat surprising that ULIRGs are seen at $z>6$
\citep[e.g.,][]{Riechers2013, Strandet2017},
as there is only $\lesssim$1~Gyr for the host galaxies to form sufficient stars
and to pollute the ISM with metals. By the same token, ULIRGs at very high
redshifts provide a new venue to probe the star formation processes in the
early universe, and therefore it is important to assemble a large sample of 
such objects and to study them systematically.

    As young stars nominally can only heat dust to a few tens of kelvin, the 
far-IR (FIR) to sub-millimeter (sub-mm) spectral energy distributions (SEDs) of
ULIRGs can be approximated well by a modified blackbody emission that has
a broad peak at roughly 80--100~$\mu$m. This offers a method to search for 
ULIRGs at high redshifts, which identifies the rising trend of the SED with
wavelength. Tailored for the Spectral and Photometric Imaging Receiver 
\citep[SPIRE;][]{Griffin2010} on board the Herschel Space Observatory
\citep[][]{Pilbratt2010}, the so-called ``500~$\mu$m peaker'' or ``500~$\mu$m 
riser'' technique selects candidates of high-redshift (hereafter ``high-$z$'')
ULIRGs by searching for sources that are progressively brighter from 250 to 500
$\mu$m \citep[][]{Pope2010, Roseboom2012}. Presumably, this method can select
ULIRGs at $z>6$ \citep[][]{Riechers2013}. Applying it to the
wide-field {\it Herschel}\, SPIRE survey data, a number of 500~$\mu$m riser
samples have been constructed \citep[][]{Dowell2014, Asboth2016, Ivison2016,
Donevski2018, Yan2020}. The method can be extended to longer wavelengths by
adding sub-mm data to the SPIRE FIR data, which could potentially select
ULIRGs at even higher redshifts. Depending on the exact location of 
the redder sub-mm band in use, it is referred to as the technique of 
``850~$\mu$m'' or ``870~$\mu$m'' riser \citep[][]{Riechers2017}. However,
the existing sub-mm surveys cover too small areas as compared to the SPIRE
surveys, and 850~$\mu$m or 870~$\mu$m risers cannot be directly selected in
a significant number. A typical sub-mm source in the existing surveys has flux 
density of a few mJy, while a typical SPIRE source has flux density of 
$\gtrsim 20$~mJy. A systematic search for 850~$\mu$m risers and the like would
have to wait for future sub-mm surveys over at least several tens of deg$^2$ in
the existing SPIRE survey fields so that a significant number of rare, bright 
(several tens of mJy) sub-mm sources would have a chance to be found in the 
first place. Nevertheless, \citet[][]{Yan2020} have attempted an alternative 
within the limit of the current 850~$\mu$m surveys done by the Submillimetre 
Common-User Bolometer Array 2 (SCUBA2). This is to search for 
``SPIRE dropouts'', which are objects prominent in 850~$\mu$m but very weak or
invisible in SPIRE and therefore could be potential 850~$\mu$m risers.

    These 500~$\mu$m risers and SPIRE dropouts are only high-$z$ candidates
and could suffer from severe contaminations by interloppers at low redshifts. 
Even for the 500~$\mu$m risers, which have well-defined colour criteria, one
would still expect a high contamination rate if the goal is to search for 
objects at $z>6$. Due to the degeneracy between dust
temperature and redshift, 500~$\mu$m risers actually can only select sources at
$z\gtrsim 4$ in general \citep[e.g.,][]{Pope2010, Yan2020}. To target the 
highest redshift range of $z>6$, \citet[][hereafter ``YMHF20'']{Yan2020} took
an innovative approach. Using the fact that high-$z$ ULIRGs should be radio 
weak, YMHF20 devised a ``high-$z$ index'' (Hi$z$Idx) to further select $z>6$
candidates among their ``Tier 1'' 500~$\mu$m risers and SPIRE dropouts. While
only a small fraction of these objects (72 out of 629 500~$\mu$m risers and 29
out of 95 SPIRE dropouts) have sufficiently deep radio data to 
allow for a meaningful assessment of Hi$z$Idx, the total number of $z>6$
candidates still amount to 20 objects (19 500~$\mu$m risers and 1 SPIRE 
dropout) and constitute the largest sample of $z>6$ ULIRG candidates.

    The true nature of such objects of course is still awaiting spectroscopic 
confirmation, which can be done in the millimeter regime with current 
technology (e.g., by identifying CO lines). On the other hand, a lot can be
learned if we can identify their counterparts in optical to
near-IR (NIR). In this regime, one would not be probing the dusty starburst 
site but would be detecting the associated host population. If the counterparts
are bright enough, one could also pursue optical-to-NIR spectroscopy to confirm
their high-$z$ nature. Moreover, one could analyze their SEDs in this regime to
gain knowledge about the host 
populations, including deriving their photometric redshifts ($z_{\rm ph}$) to 
facilitate the interpretation of their nature.

    In this work, we study the optical-to-NIR counterparts of the $z>6$ 
ULIRG candidates YMHF20 in the COSMOS field that are culled by the Hi$z$Idx
criteria. This work is enabled by the high-resolution interferometry data from 
the Atacama Large Millimeter/submillimeter Array (ALMA), which allow us to 
locate the accurate positions of our sources for follow-up studies.
We made use of the data from the
Automated Mining of the ALMA Archive in the COSMOS Field
\citep[A$^3$COSMOS][v20200310]{Liu2019}. This ongoing ALMA data mining project
processes all the non-proprietary continuum imaging data in the COSMOS field
and releases the processed images as well as the source catalogs using a number 
of different extraction methods. The rich multi-wavelength data in the COSMOS
field further enable us to study the identified sources in detail.

  Our paper is organized as follows. The sample construction is described in 
\S 2. The optical-to-NIR counterpart identification is detailed in \S 3. The
SED analysis is given in \S 4. The implication of our results is discussed in
\S 5. We conclude with a summary in \S 6. 
All magnitudes quoted are in the AB system. All coordinates are of J2000.0 
Equinox. We adopt the following cosmological parameters: 
$\Omega_M=0.27$, $\Omega_\Lambda=0.73$ and $H_0=71$~km~s$^{-1}$~Mpc$^{-1}$.

\section{Sample Description}

    Our sample is made of the objects from YMHF20 that satisfy the Hi$z$Idx 
criteria for $z>6$ and are identified in the A$^3$COSMOS data. While they are
only a very small fraction of the 500~$\mu$m risers and SPIRE dropouts in
YMHF20, these objects offer the best chance to date to study the most extreme
star formation processes in the early universe.

\subsection{ALMA identifications}

    In the COSMOS field, there are 11 500~$\mu$m risers and one SPIRE dropout 
satisfying the Hi$z$Idx criteria for $z>6$ (see the bold-faced Hi$z$Idx entries
of Table 7, 8 and 11 in YMHF20 in this field), which constitute the parent 
sample of this current work. To recapitulate YMHF20, Hi$z$Idx is the flux 
density ratio between FIR/submm and radio:
\begin{equation}\label{eq:highzindex}
    {\mathrm{Hi}}z{\mathrm{Idx}} =
    \begin{cases}
       f_{500}\times 10^{-3}/S_{1.4}\, & (500~\mu m\ {\mathrm{risers}}) \\
       S_{850}\times 10^{-3}/S_{1.4}\, & ({\mathrm{SPIRE\ dropouts}}),
    \end{cases}
\end{equation}
where $f_{500}$, $S_{850}$ and $S_{1.4}$ are the flux densities at 500~$\mu$m,
850~$\mu$m and 1.4~GHz, respectively
\footnote{To convert from the flux density at an 
arbitrary frequency to that at 1.4~GHz, YMHF20 adopted a power-law spectrum of
$S_\nu \propto \nu^{-0.8}$. For what is relevant in this study, 
$S_{1.4}=1.84\times S_{3.0}$, where $S_{3.0}$ is the flux density at 3.0~GHz.}.
Based on two known objects with spectroscopic redshifts below and above
$z=6$, respectively, YMHF20 selected $z>6$ objects using an ad hoc criterion of
Hi$z$Idx(500) $\geq 0.7$ for the 500~$\mu$m risers and Hi$z$Idx(850) $\geq 0.5$ 
for the SPIRE dropouts, respectively. The validity of these selections will be
tested in this work.

   Our current study is based on the latest source catalogs of A$^3$COSMOS
(v20200310). We used the ``blind'' catalog, which contains the sources 
extracted without using any prior information. However, the publicly available 
images are still of the previous release (v20180801). As we will show later, 
one of our identified sources does not yet have publicly available ALMA image
because of this reason. 

   While the SPIRE and the SCUBA2 data have coarse spatial resolutions
(with the beam sizes of 18\arcsec.1 and 14\arcsec.9 in SPIRE 250~$\mu$m and 
SCUBA2 850~$\mu$m, respectively), their centroid positions should still have 
1~$\sigma$ accuracy of 3\arcsec.1 and 2\arcsec.5, respectively (see YMHF20).
Here we adopted the matching radius of 9\arcsec\ when matching them to the ALMA
positions. Among the 11 high-$z$ 500~$\mu$m risers satisfying 
Hi$z$Idx(500) $\geq 0.7$, seven are in the current A$^3$COSMOS coverage and all
were matched. The only SPIRE dropout satisfying Hi$z$Idx(850) $\geq 0.5$ was 
also matched. Table 1 shows the details of these matches. As it turns out, the 
ALMA positions and the SPIRE/SCUBA2 centroid positions are all within 
$r<6.5$\arcsec.
The ALMA detections almost all have S/N $\geq 3$; the only exception is the
one for \texttt{500R\_COSMOS\_T1\_x08}, which has S/N $=2.8$. We accepted this
case because the ALMA source position coincides with that of the VLA detection
(as presented in YMHF20) almost exactly, which gives credibility of the 
identification. 

   The VLA image has the synthesized beam size of 0\arcsec.75, which is better
than most of the ALMA observations listed in Table 1. Therefore, under the
assumption that the radio and the FIR-to-mm emissions are originated from the
same region in the host, the VLA data should provide better localizations for
most of the sources. However, ALMA identifications are the least ambiguous in
our context, and we adopt the ALMA positions except in the special case of
\texttt{500R\_COSMOS\_T1\_x08}.

\begin{table*}
\centering
\tiny
\caption{ALMA identifications of Tier-1 500~$\mu$m risers and SPIRE dropout in COSMOS field
    \label{tab:almaid}}
\begin{tabular}{lccccccccc}
\hline
\hline
  Short ID       & R.A.  \&        Decl.          & f(500$|$850) & R.A.  \&    Decl.       & $S_{3.0}$ & Hi$z$Idx          &
 R.A.  \&    Decl.         & $\lambda_{\rm alma}$ & $\theta$ & $S\rm{_{alma}}$ \\
            & \multicolumn{1}{c}{(HerMES/S2CLS)}   & \multicolumn{1}{c}{(mJy)} & \multicolumn{1}{c}{(VLA)} & \multicolumn{1}{c}{($\mu$Jy)} & \multicolumn{1}{c}{(500$|$850)} &
   \multicolumn{1}{c}{(ALMA)} & \multicolumn{1}{c}{(mm)} & \multicolumn{1}{c}{(\arcsec)} & \multicolumn{1}{c}{(mJy)} \\
\hline

  500R\_x08   & 10:01:42.2   2:37:26.9 & $28.1\pm5.0$ & {\bf 10:01:42.20  2:37:27.10} &  $18.7\pm2.4$ & 0.82 & 10:01:42.17  2:37:27.47 & 1.34    & 2.76 &  $0.90\pm0.32$ \\
  500R\_x24   & 10:02:40.4   1:45:40.2 & $31.4\pm5.2$ & 10:02:40.43  1:45:44.11 &  $12.7\pm2.4$ & 1.34 & 10:02:40.44  1:45:44.17 & 0.873  & 1.27 & $10.01\pm0.37$ \\
  500R\_x26$^\dagger$   & 10:00:59.0   1:33:08.8 & $32.2\pm5.1$ & 10:00:59.18  1:33:06.73 &  $24.3\pm3.8$ & 0.72 & 10:00:59.17  1:33:06.67 & 1.03    & 0.83 & $18.36\pm0.78$ \\
              &                        &              &                         &               &      &                         & 1.31     & 1.88 &  $8.85\pm0.07$ \\
              &                        &              &                         &               &      &                         & 1.32     & 1.66 &  $9.83\pm0.08$ \\
              &                        &              &                         &               &      &                         & 1.29     & 1.20 &  $9.05\pm0.25$ \\
  500R\_x31   & 10:01:26.1   1:57:50.4 & $36.0\pm5.2$ & 10:01:26.02  1:57:51.32 &  $24.1\pm2.6$ & 0.81 & 10:01:26.03  1:57:51.27 & 0.873  & 0.84 &  $7.04\pm0.87$ \\
              &                        &              &                         &               &      &                         & 1.25     & 1.81 &  $2.24\pm0.24$ \\
\hline

  500R\_x02   & 10:01:42.3   2:00:19.3 & $26.4\pm5.0$ & 10:01:42.55  2:00:14.69 & $109.0\pm6.3$ & $>$1.25 $\rightarrow$ 0.13$^*$ & 10:01:42.54  2:00:14.67 & 0.873  & 0.85 &  $6.94\pm0.30$ \\
  500R\_x10   & 10:01:40.5   2:30:14.7 & $28.3\pm5.2$ & 10:01:40.44  2:30:10.44 &  $11.6\pm2.3$ & $>$1.34 $\rightarrow$ 1.33$^*$ & 10:01:40.43  2:30:10.55 & 1.25    & 1.79 &  $2.74\pm0.30$ \\
  500R\_x34a  & 10:00:09.9   2:22:22.0 & $39.8\pm5.1$ & 10:00:09.49  2:22:19.48 & $142.0\pm7.5$ & $>$1.87 $\rightarrow$ 0.12$^*$ & 10:00:09.49  2:22:19.47 & 0.873  & 0.29 &  $3.04\pm0.52$ \\
              &                        &              &                         &               &      &                         & 1.25    & 1.84 &  $1.06\pm0.35$ \\
  500R\_x34b  &                        &              &                         &  $40.6\pm3.1$ & $>$1.87 $\rightarrow$ 0.12$^*$ & 10:00:10.36  2:22:24.43 & 0.873  & 0.29 &  $4.98\pm0.40$ \\
              &                        &              &                         &               &      &                         & 1.25    &      &  $1.88\pm0.20$ \\

\hline\hline

  SD850\_A03  &  9:59:57.4   2:27:28.6 & $11.8\pm1.9$ &  9:59:57.29  2:27:30.54 &  $28.8\pm2.7$ & $>$0.56 $\rightarrow$ 0.22$^*$ &  9:59:57.29  2:27:30.60 & 0.873  & 0.29 & $13.87\pm0.53$ \\
              &                        &              &                         &               &      &                         & 1.25    & 1.85 &  $5.48\pm0.22$ \\
              &                        &              &                         &               &      &                         & 1.29    & 1.20 &  $4.71\pm0.24$ \\

\hline\hline
\normalsize
\end{tabular}
\tablecomments{The ``Short ID'' column omits the string ``\texttt{COSMOS\_T1\_}''
in the formal object names. The object marked with ``$\dagger$'' turns out to be
the $z=5.667$ galaxy (nickname ``CRLE'') reported in \citet[][]{Pavesi2018}. The 
Hi$z$Idx values are Hi$z$Idx(500) and Hi$z$Idx(850) for the 500~$\mu$m risers
and the SPIRE dropout, respectively; the values marked with ``*'' are the
revised values based on the new identifications of the 3~GHz counterparts, while
the corresponding values to the left side of the arrow are the original values
from Y20.
The $\theta$ values repsent the spatial resolutions of the correspondant ALMA 
observations, which are calculated by adding the beam sizes along the major 
and the minor axis in quadrature. The adopted positions are all the ALMA postions
excepted for \texttt{500R\_COSMOS\_T1\_x08}, whose position is based on the
VLA measurement (highlighted in bold-face font) because its uncertainty is
10$\times$ better than the ALMA one.
}
\end{table*}

\normalsize

\begin{figure*}
\includegraphics[height=21.5cm]{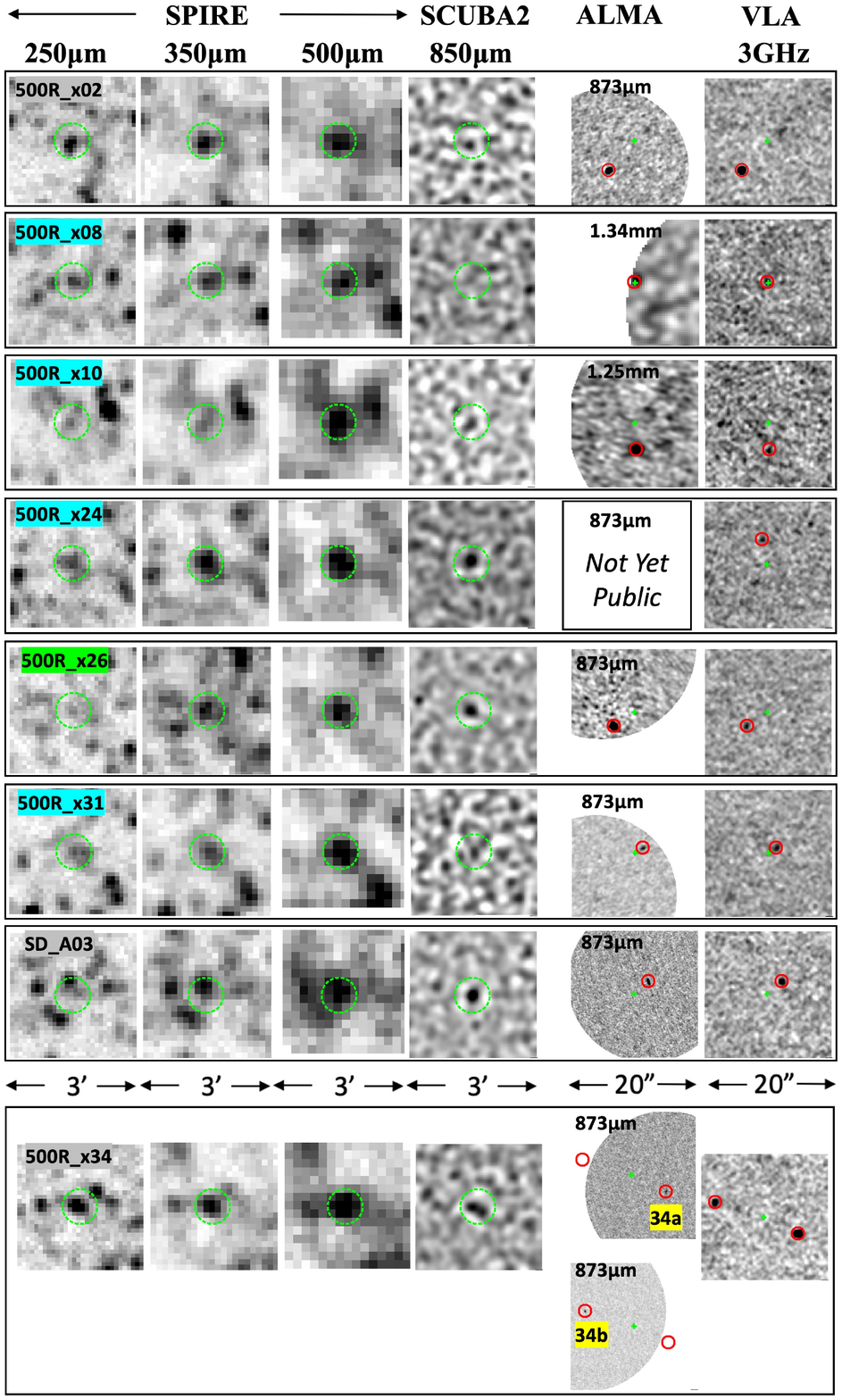}
\caption{Image stamps of the eight sources in our sample, with the source 
``Short ID'' (see Table 1) labeled in different background colours indicating
those in the high-$z$ candidate subsample (cyan or green) and those in the
contaminator subsample (grey). The SPIRE 250, 350, 500~$\mu$m and the SCUBA2 
850~$\mu$m images are 3\arcmin$\times$3\arcmin\ in size, and the dashed green
circles (25\arcsec\ in radius) are centred on the SPIRE (for 500~$\mu$m
risers) or the SCUBA2 (for SPIRE dropout) positions. The ALMA identification
images (with the central wavelengths of the identifications labeled) and the
VLA images are 20\arcsec$\times$20\arcsec\ in size; the red circles
(1\arcsec\ in radius) are centred on the ALMA positions, while the green 
crosses mark the SPIRE/SCUBA2 positions. \texttt{500R\_COSMOS\_T1\_x24} does
not yet have the ALMA images publicly available. \texttt{500R\_COSMOS\_T1\_x34}
has two ALMA counterparts (``\texttt{a}'' and ``\texttt{b}''), which are 
labeled as ``34a'' and ``34b'' in yellow background colour. 
}
\label{fig:ID_ALMA_VLA}
\end{figure*}

    Figure \ref{fig:ID_ALMA_VLA} shows the SPIRE and the SCUBA2 images for all
these eight objects as well as their ALMA identification images and their 
images in the VLA 3~GHz. In particular, \texttt{500R\_COSMOS\_T1\_x34} is 
identified with two ALMA sources, which we call the ``\texttt{a}'' and 
``\texttt{b}'' components, respectively.

\subsection{Detections and non-detections in the VLA Image}

    We first revisit the Hi$z$Idx assessment of these objects based on the new
identifications, because the calculations of Hi$z$Idx in YMHF20 used the 
${\rm S/N=5}$ upper limit of the VLA 3~GHz map for the non-detections in this
map. YMHF20 matched the 500~$\mu$m risers and the SPIRE dropouts with the VLA 
sources using the criterion of $\Delta/\sigma\leq 1.5$, where $\Delta$ is the 
positional offset to the nearest 3~GHz source and $\sigma$ is the overall 
positional uncertainty by adding in quadrature the errors of the reported 
SPIRE (or SCUBA2) and VLA positions. $\sigma$ is dominated by the SPIRE (or 
SCUBA2) positional error, which depends on the source S/N in the SPIRE/SCUBA2
data, and the error term from the VLA positions is minimal (set to
0\arcsec.13 at the fixed S/N $= 5$). For the 500~$\mu$m risers and the SPIRE
dropouts, $\sigma\leq$~3\arcsec.1 and 2\arcsec.2, respectively.

    Following this matching criterion, four of the 11 aforementioned 500~$\mu$m 
risers were labeled as detected in radio by YMHF20. These are 
\texttt{500R\_COSMOS\_T1\_x08}, \texttt{x24}, \texttt{x26}, and \texttt{x31}, 
all among the seven that have ALMA coverage. The ALMA counterparts very well 
coincide with the VLA counterparts, and the slight differences in the 
positions are all consistent with the respective positional errors. Their
Hi$z$Idx values remain the same as reported in YMHF20. Seven of the 11 sources 
were labeled by YMHF20 as undetected in radio, and their reported Hi$z$Idx 
values were above the threshold because of the use of the ${\rm S/N=5}$
upper limit for $S_{3.0}$. 
On the other hand, YMHF20 did carry out a more generous 
match using a searching radius of 10\arcsec, and found that three of these 
seven had corresponding 3~GHz matches; they were still labeled as radio
non-detections in YMHF20 only because these matches have $\Delta/\sigma>1.5$
(see the bottom part of Table 7 in YMHF20). These three are 
\texttt{500R\_COSMOS\_T1\_x02}, \texttt{x04} and \texttt{x10}, which are also
within the ALMA coverage. The ALMA identifications now show that these 3~GHz 
matches are in fact the right counterparts (i.e., the no-match criterion in
YMHF20 should be relaxed to $\Delta/\sigma>3$). Therefore, their Hi$z$Idx must
be revised. Similarly, the only previous $z>6$ candidate among the 
SPIRE dropouts, \texttt{SD850\_COSMOS\_T1\_A03}, is of the same situation
(see the bottom part of Table 10 in YMHF20)
\footnote{We note that Table 10 in YMHF20 (for the radio matching of the SPIRE
dropouts) has an error in reporting the $\Delta_{\rm pos}$ and 
$\sigma_{\rm pos}$ values for the bottom six objects; the values in these two 
columns should be swapped.} and its Hi$z$Idx must also be revised.

   The updated Hi$z$Idx(500$|$850) values are included in Table 1. To 
summarize, five of the seven 500~$\mu$m risers satisfy Hi$z$Idx(500) $\geq 0.7$
and are candidates at $z>6$ (including \texttt{500R\_COSMOS\_T1\_x10} whose 
Hi$z$Idx(500) has been revised), which we will refer to as the ``high-$z$ 
candidates'' in the rest of this paper. The other two 500~$\mu$m risers 
(including \texttt{500R\_COSMOS\_T1\_x34}, which is still considered
as a single source in this context) no longer satisfy this criterion, which
we will refer to as the ``contaminators'' hereafter. The SPIRE dropout,
\texttt{SD850\_COSMOS\_T1\_A03}, is also a contaminator. These eight sources
constitute the sample of this study.

\section{Multi-wavelength properties of the counterparts}

\subsection{Data Description}

   The COSMOS field has accumulated a wealth of multi-wavelength data, which
make it possible to further study the properties our sources. These
include the {\it HST}\ Wide Field Camera 3 (WFC3) F160W (hereafter $H_{160}$)
images from the COSMOS-Drift And SHift program \citep[COSMOSDASH;][]{Mowla2019}
and Advanced Camera for Surveys (ACS) F814W (hereafter $I_{814}$) image that 
defines the original COSMOS field \citep[][]{Koekemoer2007}, the VISTA 
VIRCAM $Y$, $J$, $H$, and $K_s$ images from the UltraVISTA program
\citep[][]{McCracken2012}, the Subaru Hyper Suprime-Cam $g$, $r$, $i$, $z$, and
$y$ images from the Hyper Suprime-Cam Subaru Strategic Program 
\citep[HSC-SSP;][]{Aihara2018a,Aihara2018b}, and the {\it Spitzer} Infrared 
Array Camera (IRAC) and Multi-Band Imaging Photometer for Spitzer (MIPS) images
from the Spitzer COSMOS program \citep[S-COSMOS;][]{Sanders2007}. 
These datasets are briefly described below.
In most cases, we had to carry out our own photometry using SExtractor
\citep[][]{Bertin1996}, for which we adopted its \texttt{MAG\_AUTO}. In some 
cases we adopted the photometry from the relevant data releases, and these
cases will be explictly noted in the rest of the paper.

   {\it HST WFC3 $H_{160}$.}\  We used the v1.2.10 data released by the 
COSMOS-DASH program, which incorporate its own observations as well as those 
from other programs in this field. The image mosaic effectively covers 0.66 
deg$^2$ and has the scale of 0\arcsec.1~pix$^{-1}$. The full-width at 
half-maximum (FWHM) of the point spread function (PSF) is 0\arcsec.21. The 
mosaic has reached the 5~$\sigma$ sensitivity of $AB = 25.1$~mag for point
sources (using an aperture of 0\arcsec.3 in diameter) in general, but in some
regions it is significantly deeper. 

   {\it HST ACS $I_{814}$.}\  We used the image mosaic with the scale of
0\arcsec.03~pix$^{-1}$ included in the v2.0 data release of the COSMOS
program. The image covers $\sim$1.7~deg$^2$ and reaches the 5~$\sigma$
sensitivity of $AB=27.2$~mag for point sources sources (0\arcsec.24 diameter
aperture). The PSF FWHM is 0\arcsec.095.

   {\it VISTA near-IR.}\ We used the v4.0 $YJHK_s$ image stacks of the
UltraVISTA program, which cover $1.5\times 1.2$~deg$^2$ in four ``Ultra-deep''
and four ``Deep'' stripes. The PSFs in all these four bands have FWHM of
0\arcsec.76--0\arcsec.78. The 5~$\sigma$ limiting magnitudes (2\arcsec\ diameter
aperture) in $Y$, $J$, $H$, and $K_s$ are 25.8, 25.6, 25.2, and 24.9~mag
in the Ultra-deep stripes, respectively, and 24.7, 24.5, 24.1, and 24.5~mag in
the Deep stripes, respectively.

   {\it Subaru optical.}\ We used the five broad-band image stacks from the 
third public data release (PDR3) of the HSC-SSP program, which has observed
the COSMOS field as one of the survey's
UltraDeep fields. These data have reached the 5~$\sigma$ limits (within
aperture of 2\arcsec\ diameter) of 28.2, 27.9, 27.7, 27.1, and 26.1~mag in $g$,
$r$, $i$, $z$, and $y$, respectively, with the PSF sizes varying from 
0\arcsec.7 to 0\arcsec.8.

   {\it Spitzer IR.}\ We used the data from the S-COSMOS program, which include
the IRAC data in Ch1 through Ch4 (3.6 to 8.0~$\mu$m) and the MIPS data in 
24~$\mu$m. For the IRAC photometry, we adopted the GO2 catalog. Specifically,
we used the fluxes extracted through a 2\arcsec.9 aperture. These fluxes were 
corrected to the total fluxes by dividing the correction factors of 0.90, 0.90,
0.84, and 0.73 in Ch1, Ch2, Ch3, and Ch4, respectively, as per the instruction
of the data release. The MIPS data were only used to aid the visual inspection
of the sources; for the sake of completeness, the 24~$\mu$m photometry, which
was taken from the GO3 catalog, will be listed when appropriate.

\begin{table*}
\centering
\small
\caption{Photometric Information
    \label{tab:phot}}
\begin{tabular}{ccccccccc}
\hline\hline
  Short ID        & 500R\_x08$^*$  & 500R\_x24$^*$  & 500R\_x31$^*$  & 500R\_x10$^*$  &    500R\_x02   &   500R\_x34a   &   500R\_x34b   &   SD850\_A03   \\
\hline
    $g$           &     $>$28.2    & ?28.7$\pm$0.4  &     $>$28.2    &     $>$28.2    & 26.78$\pm$0.07 & 25.03$\pm$0.05 & 27.14$\pm$0.20 & 27.23$\pm$0.10 \\
    $r$           &     $>$27.9    &     $>$27.9    &     $>$27.9    &     $>$27.9    & 25.82$\pm$0.05 & 24.13$\pm$0.03 & 26.64$\pm$0.19 & 26.78$\pm$0.10 \\
    $i$           &     $>$27.7    &     $>$27.7    &     $>$27.7    &     $>$27.7    & 25.51$\pm$0.04 & 23.77$\pm$0.03 & 25.89$\pm$0.10 & 25.77$\pm$0.05 \\
    $z$           &     $>$27.1    &     $>$27.1    &     $>$27.1    &     $>$27.1    & 25.02$\pm$0.05 & 23.23$\pm$0.03 & 25.33$\pm$0.11 & 25.27$\pm$0.05 \\
    $y$           &     $>$26.1    &     $>$26.1    &     $>$26.1    &     $>$26.1    & 25.00$\pm$0.08 & 24.15$\pm$0.13 & 25.44$\pm$0.22 & 24.91$\pm$0.07 \\
\hline
    $Y$           &     $>$25.8    &     $>$24.7    &     $>$24.7    &     $>$25.8    & 24.73$\pm$0.04 & 22.35$\pm$0.01 & 24.10$\pm$0.03 &     $>$24.7    \\
    $J$           & 22.47$\pm$0.04 &     $>$24.5    & 24.91$\pm$0.31 & 26.52$\pm$0.81 & 24.31$\pm$0.03 & 21.57$\pm$0.01 & 23.97$\pm$0.03 &     $>$24.5    \\
    $H$           & 21.61$\pm$0.02 &     $>$24.1    & 24.46$\pm$0.24 & 24.53$\pm$0.17 & 23.29$\pm$0.02 & 21.30$\pm$0.01 & 23.01$\pm$0.02 &     $>$24.1    \\
    $K_s$         & 21.03$\pm$0.02 &     $>$24.5    & 24.19$\pm$0.14 & 23.55$\pm$0.09 & 22.93$\pm$0.02 & 20.89$\pm$0.01 & 22.24$\pm$0.01 & 23.77$\pm$0.16 \\
\hline
  $I_{814}$       &     $>$27.2    &     $>$27.2    &     $>$27.2    &     $>$27.2    &     $>$27.2    & 26.67$\pm$0.22 &     $>$27.2    &     $>$27.2    \\
  $H_{160}$       & 22.15$\pm$0.11 &       N/A      & 24.61$\pm$0.25 &     $>$25.1    & 24.02$\pm$0.29 & 21.60$\pm$0.09 & 23.35$\pm$0.33 &       N/A      \\
\hline
 Ch1 (3.6~$\mu$m) &      ...       &     $>$24.1    & 22.35$\pm$0.05 &      ...       & 21.72$\pm$0.03 & 20.32$\pm$0.01 & 20.70$\pm$0.01 & 22.88$\pm$0.07 \\
 Ch2 (4.5~$\mu$m) &      ...       &     $>$23.3    & 21.75$\pm$0.04 &      ...       & 21.32$\pm$0.03 & 20.02$\pm$0.01 & 20.31$\pm$0.01 & 22.69$\pm$0.10 \\
 Ch3 (5.8~$\mu$m) &      ...       &     $>$21.3    & 21.40$\pm$0.13 &      ...       & 21.04$\pm$0.10 & 19.92$\pm$0.03 & 20.04$\pm$0.04 &     $>$21.3    \\
 Ch4 (8.0~$\mu$m) &      ...       &     $>$21.0    & 21.52$\pm$0.33 &      ...       & 20.55$\pm$0.13 & 20.36$\pm$0.11 & 20.49$\pm$0.13 &     $>$21.0    \\
\hline
 MIPS (24~$\mu$m) &      ...       &     $<$71      &     $<$71      &      ...       &   167$\pm$16   &   381$\pm$14   &   414$\pm$15   &     $<$71      \\
\hline\hline
\normalsize
\end{tabular}
\tablecomments{The ``Short ID'' row is the same as in Table 1. The objects
labeled with ``$^*$'' are high-$z$ candidates based on the revised Hi$z$Idx,
while those without are contaminators. Object \texttt{500R\_x26} is not 
included for the reason explained in the text.  From top to bottom, the 
photometry are based on the data from the Subaru HSC-SSP in $grizy$, the 
UltraVISTA in $YJHK_s$, the COSMOS {\it HST}\ program in ACS $I_{814}$, the 
COSMOS-DASH in WFC3 $H_{160}$, the S-COSMOS in IRAC Ch1 through Ch4 and in MIPS 
24~$\mu$m. The photometry is quoted in AB magnitudes except for MIPS 24~$\mu$m,
where the values are in $\mu$Jy. The limits are 5~$\sigma$ limits, while no
values (``...'') indicate that the sources are too severely blended to
extract fluxes in these bands. The
$g$-band magnitude for \texttt{500R\_x24} is marked with ``?'', indicating that
the ``detection'' could be a false positive. 
}
\end{table*}

\subsection{Counterparts of the high-$z$ candidates}

    We start from identifying the optical-to-NIR counterparts of the
five high-$z$ candidates using the positions verified by the ALMA data.

\begin{figure*}
\includegraphics[width=14cm]{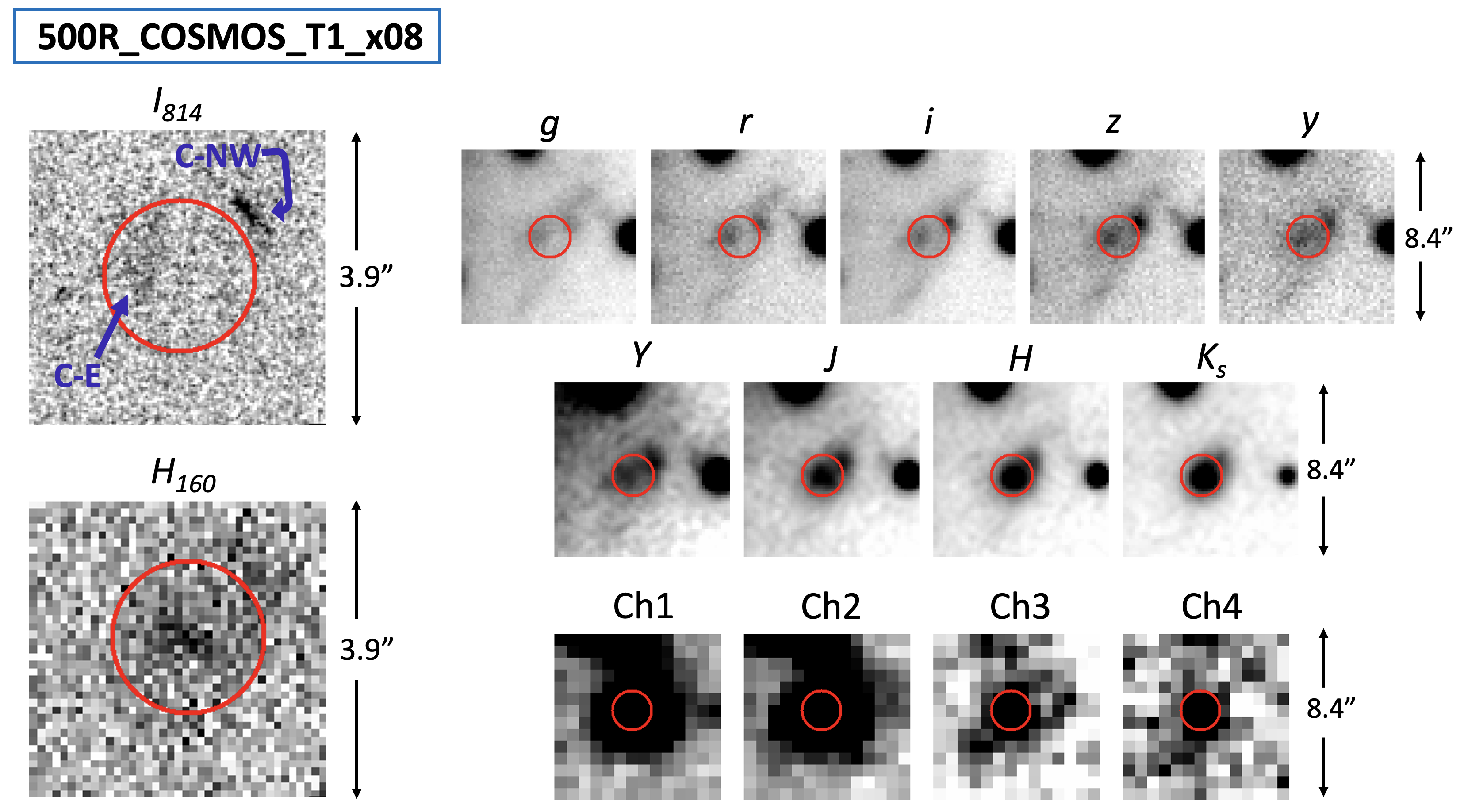}
\caption{Image stamps showing the optical-to-NIR counterpart of
\texttt{500R\_COSMOS\_T1\_x08} and its vicinity. The red circles are 1\arcsec\
in radius, and are centred on the VLA position as reported in Table 1.
Passbands are labeled. The left panel shows the high-resolution {\it HST}\
images in ACS $I_{814}$ (top) and WFC3 $H_{160}$ (bottom), which are 
3\arcsec.9$\times$3\arcsec.9\ in size. The counterpart is clearly detected in
$H_{160}$ but is invisible in $I_{814}$. Two companion objects by projection,
``C-NW'' and ``C-E'', are visible in $I_{814}$. The right panel shows the 
images in the HSC $g$, $r$, $i$, $z$, and $y$ (top), the VISTA $Y$, $J$, $H$,
and $K_s$ (middle), and the IRAC Ch1, Ch2, Ch3, and Ch4 (bottom), all 
8\arcsec.4$\times$8\arcsec.4 in size. The counterpart is not visible in $griz$
but is clearly detected in $JHK_s$ images. The companion C-E is visible in from
$g$ to $Y$, and C-NW persists through $K_s$. C-NW is severely blended with the
counterpart in the IRAC images.
}
\label{fig_500R_x08}
\end{figure*}

\begin{figure*}
\includegraphics[width=14cm]{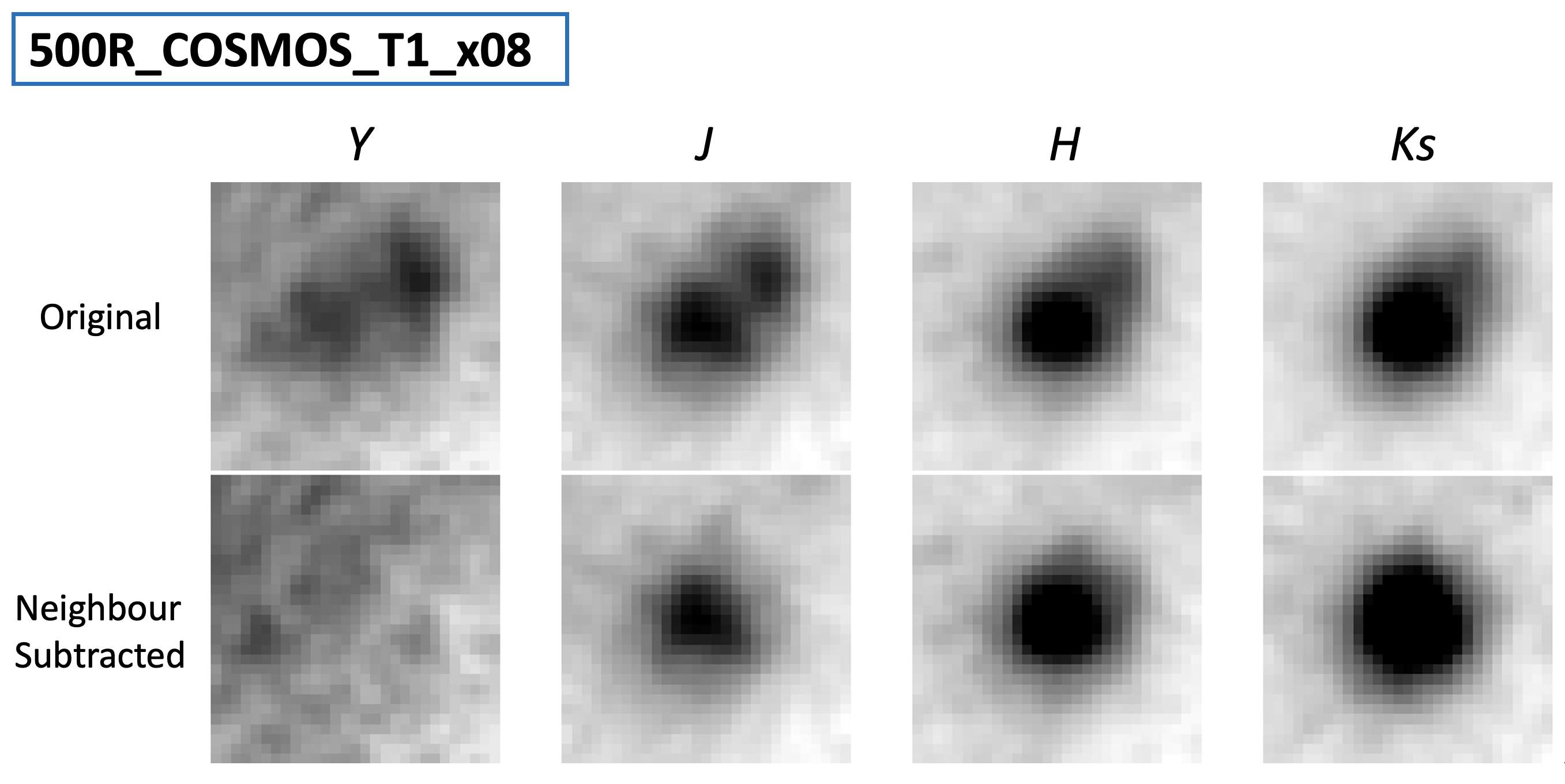}
\caption{
Subtraction of the close neighbours to \texttt{500R\_COSMOS\_T1\_x08}
in the UltraVISTA $YJHK_s$ images. The subtraction is explained in the text.
The image sizes are all 4\arcsec.35$\times$4\arcsec.35 in size and are oriented
north-up/east-left.
The top row shows the original images centred on the target, while the bottom 
row shows the neighbour-subtracted, ``cleaned'' images. The companions 
C-NW and C-E (see Figure \ref{fig_500R_x08}) are subtracted from $Y$, and the
counterpart is not visible. As C-E is no longer visible in $JHK_s$, only
C-NW is subtracted from these images.
}
\label{fig_x08_sub}
\end{figure*}

\subsubsection{500R\_COSMOS\_T1\_x08}

    As mentioned earlier, its ALMA detection has only S/N $=2.8$. This is 
largely because the source is at the very edge of the field. However the 
separation of ALMA and radio position is only 0\arcsec.58, and this lends 
strong support to the identification and the localization of the source. We 
adopt the VLA position as its final position, as the ALMA position has much 
larger uncertainty of 0\arcsec.85 (as compared to the uncertainty of 
0\arcsec.08 of the VLA position). The optical-to-NIR images of this source and 
its vicinity are shown in Figure \ref{fig_500R_x08}. At this position, there is
no visible source in $I_{814}$, and only a diffuse source in $H_{160}$ of 
$22.15\pm0.11$~mag by our photometry. This source has two very
close companions at $\sim$1\arcsec, both of which could confuse the 
identification and contaminate its photometry. We designate the one to its
north-west as ``C-NW'' and the one to its east as ``C-E'', respectively.
In $I_{814}$, C-NW is clearly an edge-on disk galaxy ($I_{814}=25.12\pm0.13$
mag), while C-E is very diffuse and is extracted as three components 
($I_{814}=25.67\pm0.18$~mag after combing the three).
In $H_{160}$, C-E is not visible and C-NW is barely detected 
($H_{160}=22.87\pm0.22$~mag).

   This source is invisible in the HSC
$griz$ images. There is a hint that it is detected in $y$, however, it is 
severely blended with the companion C-E, which becomes fainter in the redder
bands but is still visible in $y$. This is evident from our own extraction
of C-E in $y$, whose centroid has an offset of 0\arcsec.3 with respect to
that extracted in $g$ band. The offset is towards the position of
\texttt{500R\_x08}, which suggests that it is caused by the blending of 
the weak detection of this 500~$\mu$m riser in $y$. The photometry in $y$ is
further complicated by the contamination from the edge of the bright halo of 
the saturated star in its vicinity. The UltraVISTA images provide more clues,
although \texttt{500R\_x08} and the nearby objects are not included in the 
released catalog (presumably due to the contamination of the saturated star). 
The blending with C-E is more clearly seen in the UltraVISTA $Y$-band. The 
500~$\mu$m riser becomes distinctly visible in the UltraVISTA $J$-band, and is
brighter in $H$ and $K_s$. C-E seems to disappear in $JHK_s$. C-NW, on
the other hand, persists from $g$ through $K_s$. 
Nevertheless, our extraction can still separate C-NW from the 
500~$\mu$m riser in $JHK_s$. 
Therefore, we used \texttt{GALFIT}
\citep[][]{Peng2002, Peng2010} to fit C-NW and to
subtract its best-fit S\'{e}rsic profile from the $JHK_s$ 
images for better photometry. For the $Y$-band image, we further subtracted 
C-E using the same procedure and found that the 500~$\mu$m riser was not 
visible in this residual image. The comparison of the original and the 
neighbour-subtracted $YJHK_s$ images are shown in Figure \ref{fig_x08_sub}. 
The final photometry of the 500~$\mu$m riser counterpart was done on 
these ``cleaned'' images using SExtractor in the dual-image mode, where the
$K_s$ image was chosen as the detection image to define the 
\texttt{``MAG\_AUTO''} apreture. The results are listed in Table 2.
Unfortunately, C-NW cannot be separated 
from the 500~$\mu$m riser in the IRAC images, and we refrain from obtaining
photometry of the 500~$\mu$m riser on these images.


\begin{figure*}
\includegraphics[width=14cm]{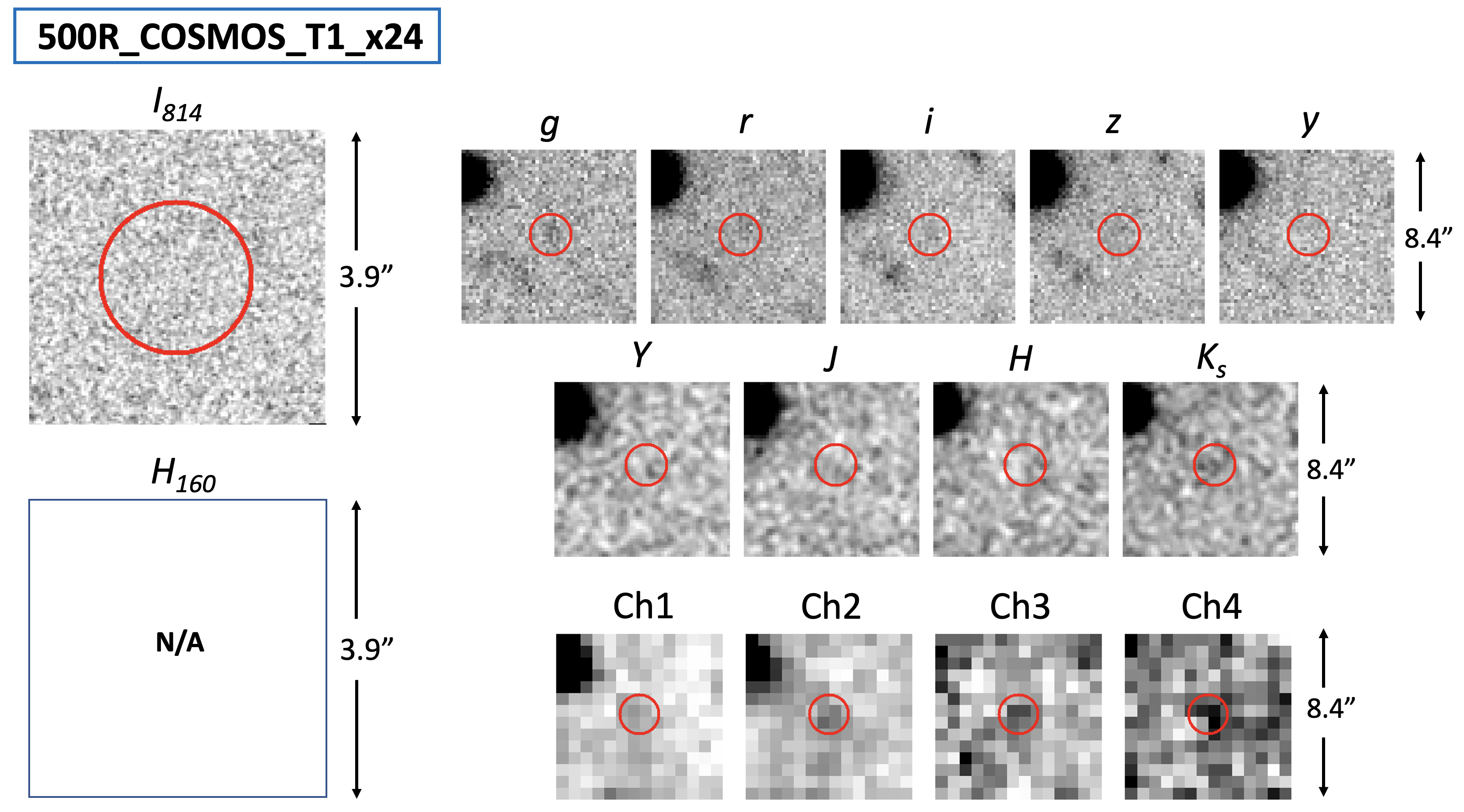}
\caption{Optical-to-NIR images at the vicinity of
\texttt{500R\_COSMOS\_T1\_x24}. The image arrangement is the same as in 
Figure \ref{fig_500R_x08}, except that there is no $H_{160}$ image available. 
The red circles ($r=1$\arcsec) are centred on the ALMA position. The 
counterpart is invisible in $I_{814}$. It is also invisible in all bands except
in the HSC $g$ band, where there is a very weak positive signal
(S/N $\approx 2.5$) at its location. However, this ``signal'' could also be a 
false positive. The seemingly positive ``detection'' in IRAC Ch4 is consistent 
with being a noise spike.
}
\label{fig_500R_x24}
\end{figure*}

\subsubsection{500R\_COSMOS\_T1\_x24}

  While the A$^3$COSMOS image for this 
source is not yet available, the A$^3$COSMOS position coincides with that of
the VLA position (0\arcsec.16 separation). Figure \ref{fig_500R_x24} shows the
optical-to-NIR images around this location. There is no visible $I_{814}$
source at the ALMA position. Unfortunately, it has no $H_{160}$ image because
it is out of the COSMOS-DASH coverage.

    This source is invisible in the HSC $rizy$ images. On the other hand, there 
seems to be a weak detection in $g$. In our extraction, it has 28.7$\pm$0.4~mag
at the exact location. However, due to its low S/N, it is highly likely that
this ``detection'' is a false positive. The source is also invisible in the 
UltraVISTA images and the S-COSMOS IRAC images. We cannot rule out the 
possibility that it is at high-$z$.

\subsubsection{500R\_COSMOS\_T1\_x26}

   This object in fact is the $z=5.667$
galaxy reported by \citet[][]{Pavesi2018}. Nicknamed ``CRLE'', it was 
serendipitously discovered during ALMA observations of a normal galaxy in its
vicinity, and its redshift was determined through the identification of 
atomic fine-structure lines ([C \RN{2}] 158~$\mu$m and [N \RN{2}] 205~$\mu$m)
in the ALMA data as well as the CO(2-1) 1.301~mm transition line in the
follow-up VLA observations. Had there not been such spectroscopic
identifications, it would be impossible to reveal the nature of this source
because it is completely blocked by a foreground, edge-on disk galaxy 
at $z_{\rm ph} \sim 0.35$ from UV to mid-IR; its FIR-to-mm emissions,
on the other hand, are transmissible through this foreground galaxy
\citep[][]{Pavesi2018}. Its redshift, however, is still below $z=6$. We
refer the reader to \citet[][]{Pavesi2018} for the optical-to-NIR images
around this position, and will not discuss this source further because
the optical-to-NIR images would provide information only on the foreground
object.

\begin{figure*}
\includegraphics[width=14cm]{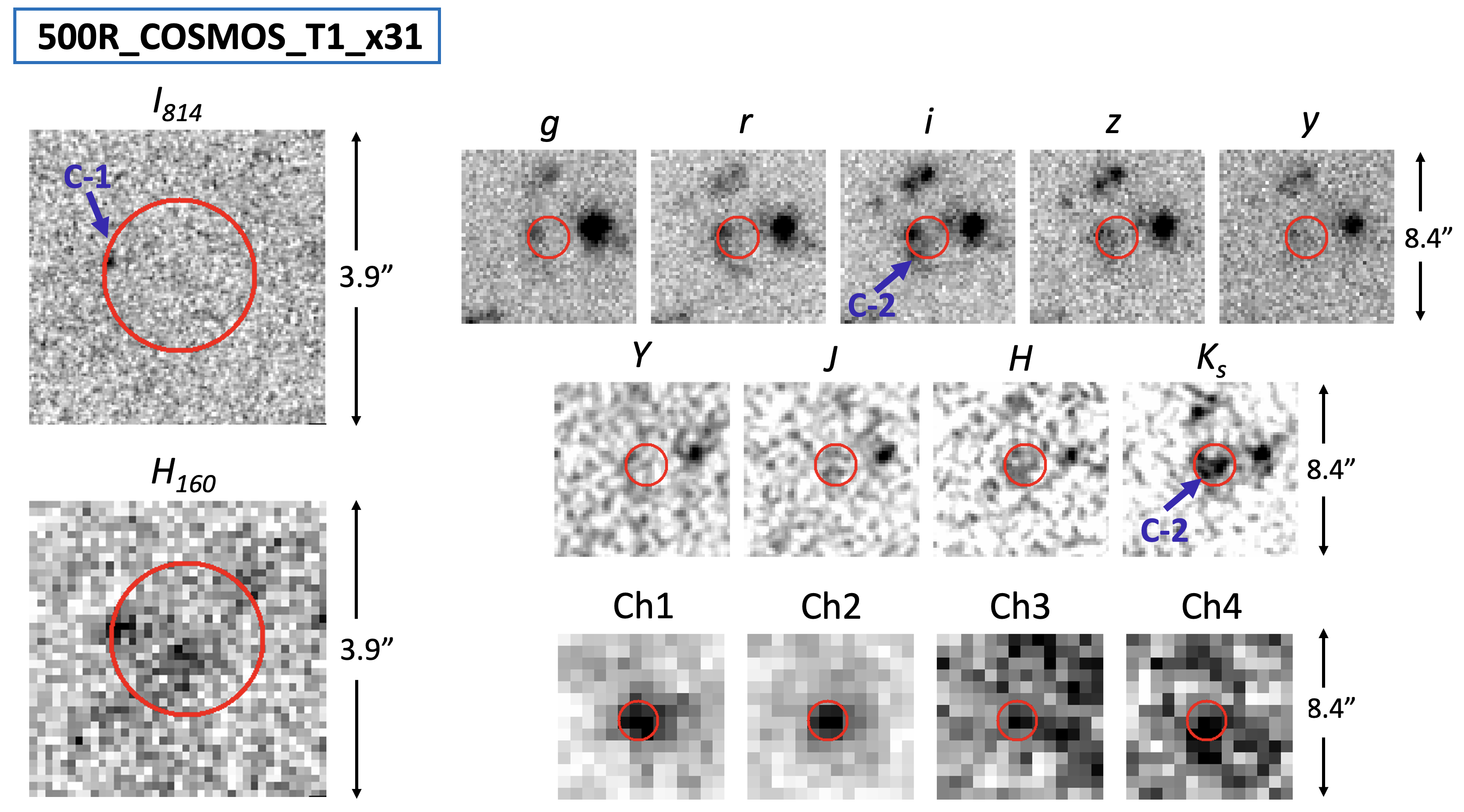}
\caption{Image stamps showing the optical-to-NIR counterpart of
\texttt{500R\_COSMOS\_T1\_x31} and its vicinity. The image arrangement is the
same as in Figure \ref{fig_500R_x08}, and the red circles ($r=1$\arcsec) are
centred on the ALMA position. The counterpart is invisible in $I_{814}$, but
appears in $H_{160}$ as a faint detection. A companion object by projection, 
labeled as ``C-1'' in $I_{814}$, is visible in both $I_{814}$ and $H_{160}$. The
counterpart is clearly detected in the VISTA $K_s$ and barely visible in the
VISTA $H$, but is not visible in any bluer bands. It has a second companion
object by projection, ``C-2'', which is the most obviously seen in the HSC $i$
and the VISTA $K_s$. The counterpart is detected in IRAC Ch1 and Ch2, but it
might be blended with the two companions if they persist in these two bands.
Its IRAC Ch3 and Ch4 images are consistent with being noise.
}
\label{fig_500R_x31}
\end{figure*}

\begin{figure*}
\includegraphics[width=14cm]{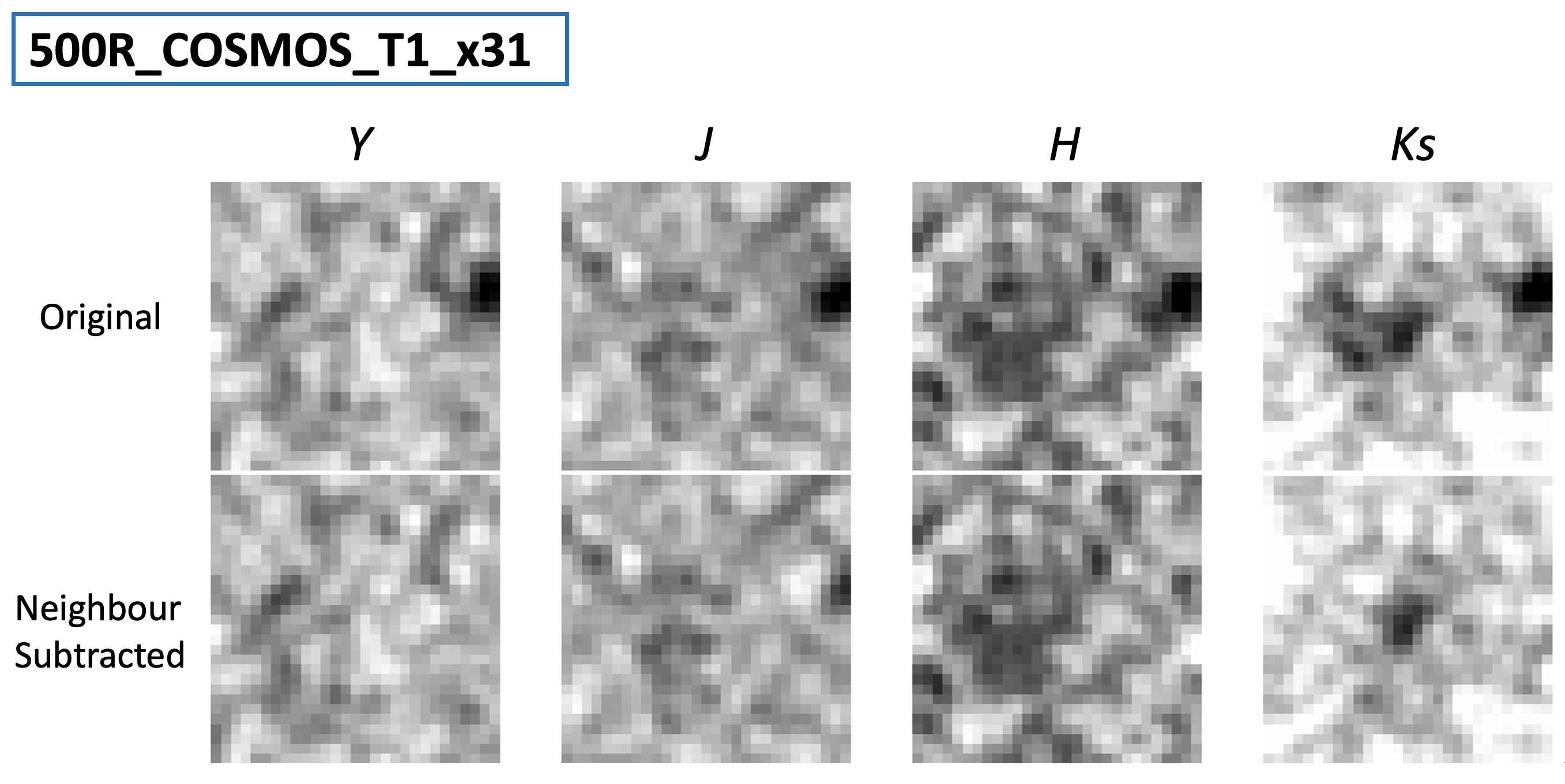}
\caption{
Same as Figure \ref{fig_x08_sub}, but for 
\texttt{500R\_COSMOS\_T1\_x31}. The two close neighbours, C-1 and C-2 (see
Figure \ref{fig_500R_x31}), are almost invisible in the original $YJH$ images,
and therefore their subtractions do not create any notable effect in these 
three bands.
}
\label{fig_x31_sub}
\end{figure*}

\subsubsection{500R\_COSMOS\_T1\_x31}

   The optical-to-NIR images of this source are shown in 
Figure \ref{fig_500R_x31}. At the exact ALMA location, there is a faint
blob in $H_{160}$, which should be the counterpart. By our own photometry, it 
has $H_{160}$$=$24.61$\pm$0.25~mag. It is invisible in $I_{814}$.  
There is a close companion
$\sim$0.\arcsec 77 away (R.A. $=$ 10:01:26.08, Decl. $=$ 1:57:51.43), which
is detected in both bands. We designate this source as ``C-1''. In the COSMOS 
catalog, this companion
has $I_{814}=25.72\pm0.14$ and $H_{160}=25.65\pm0.11$~mag, respectively. 

    C-1 is clearly visible in all the HSC images. A second close neighbour, C-2,
is also revealed by the HSC images in $r$ and $i$, which is 1\arcsec.04 away 
from the ALMA centroid.  Using the $i$-band image
for detection, our run of SExtractor in dual-image mode gives photometry for
C-1 and C-2 in $grizy$ as 26.93$\pm$0.21, 25.81$\pm$0.12, 25.52$\pm$0.08, 
25.60$\pm$0.17, 25.78$\pm$0.40~mag and
27.42$\pm$0.31, 26.06$\pm$0.14, 25.82$\pm$0.12, 25.80$\pm$0.20, 25.88$\pm$0.44
mag, respectively. The true counterpart is revealed in the UltraVISTA images,
especially in $K_s$. The UltraVISTA DR4 catalog, however, only extracts this 
counterpart together with C-1 and C-2 as a single, blended object.
We were able to separate the three in $K_s$. Following the same 
procedure as in the case of \texttt{500R\_COSMOS\_T1\_x08}, we subtracted C-1
and C-2 from the UltraVISTA images for the photometry of the 500~$\mu$m riser
counterpart. The comparison of the original and the neighbour-subtracted images
is shown in Figure \ref{fig_x31_sub}. The subtraction in $YJH$ does not create 
any notable effect, as C-1 and C-2 are almost invisible in these bands.
The $YJHK_s$ \texttt{MAG\_AUTO} photometry, done using the neighbour-subtracted
$K_s$ image as the detection image to define the aperture, are listed in 
Table 2.

In the S-COSMOS catalog, there is an IRAC source at this 
location, and the results are listed in Table 2. From the IRAC images, one 
could argue that C-1 and C-2 might not be visible and that the fluxes are all 
contributed by the 500~$\mu$m riser counterpart. However, this is admittedly
uncertain, and we will consider two scenarios, with and without the IRAC 
photometry, respectively, in the follow-up SED analysis.

\begin{figure*}
\includegraphics[width=14cm]{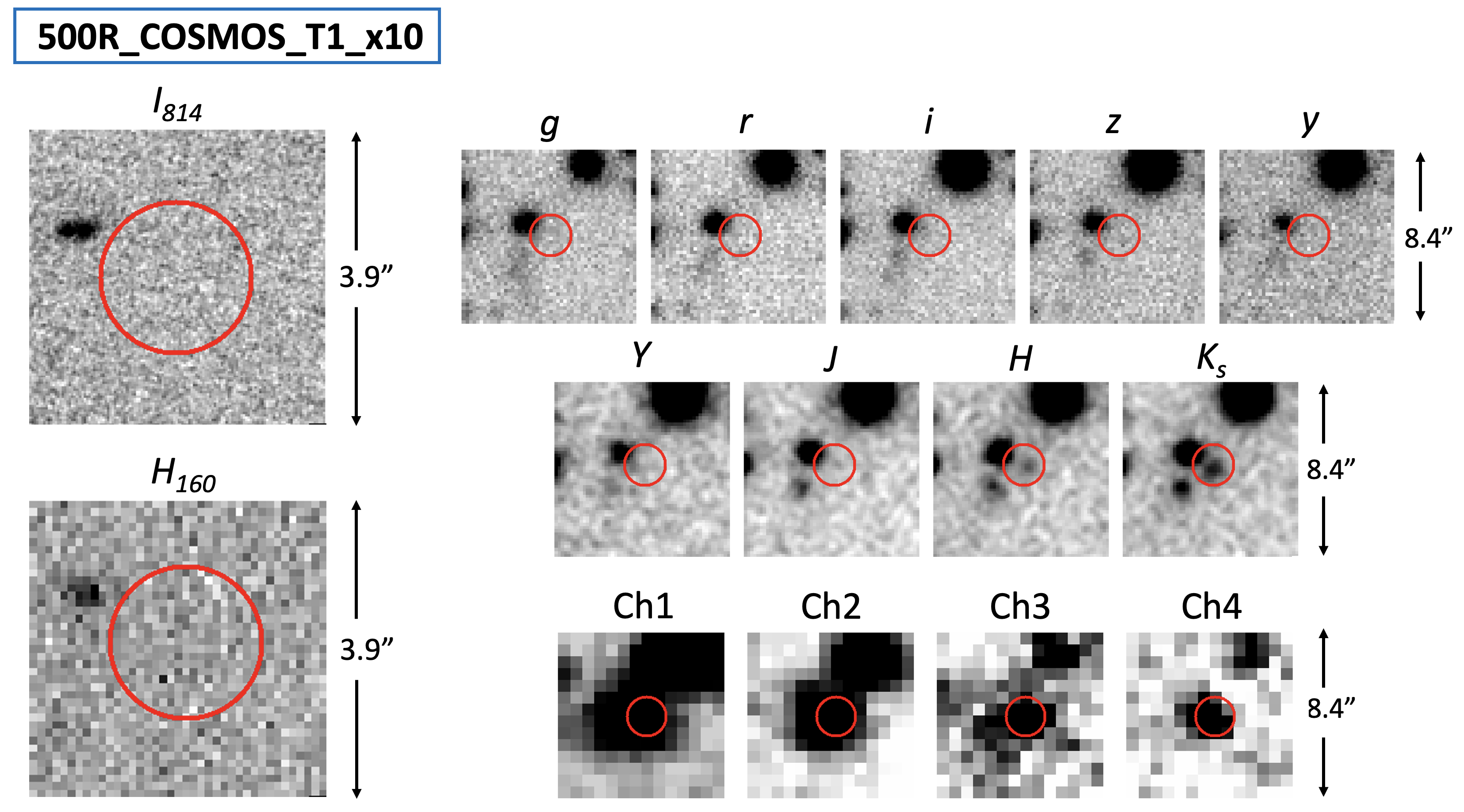}
\caption{Image stamps showing the optical-to-NIR counterpart of
\texttt{500R\_COSMOS\_T1\_x10} and its vicinity. The image arrangement is the
same as in Figure \ref{fig_500R_x08}, and the red circles ($r=1$\arcsec) are 
centred on the ALMA position. The counterpart is invisible in either $I_{814}$
or $H_{160}$. It is prominent in the VISTA $K_s$ and $H$ images and barely
visible in $J$, but is invisible in any bluer bands. Its $K_s$ image appears
bigger than in $H$ and $J$; in other words, it seems that only part of its 
$K_s$ counterpart is visible in $H$ and $J$. There are two unrelated objects
in the neighbourhood,
which are more than 1\arcsec away and are not labeled. The counterpart is
visible in all IRAC channels, however it is severely blended with at least one
of the neighbours.
}
\label{fig_500R_x10}
\end{figure*}

\begin{figure*}
\includegraphics[width=14cm]{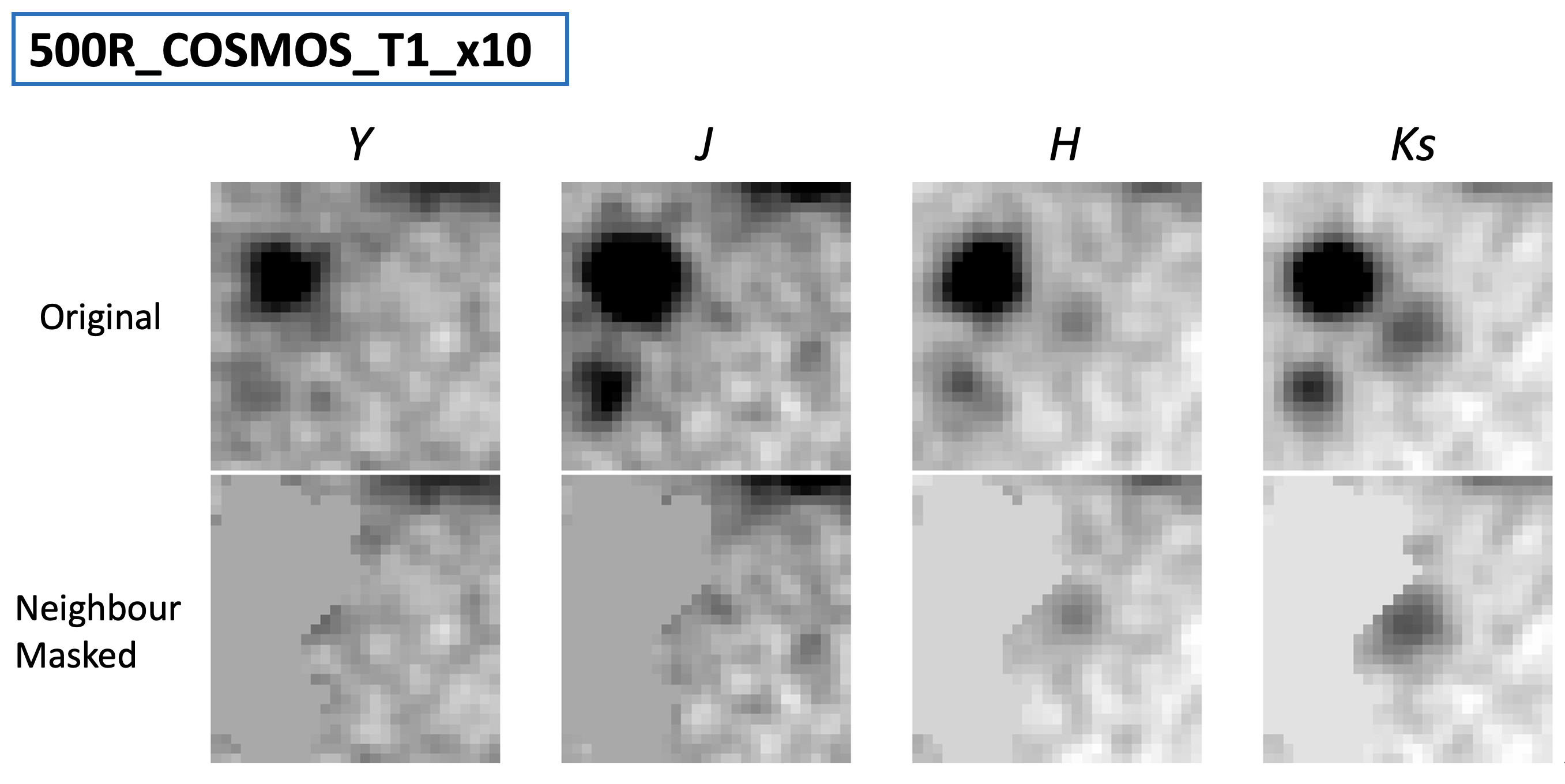}
\caption{
Similar to Figures \ref{fig_x08_sub} and 
\ref{fig_x31_sub}, but for \texttt{500R\_COSMOS\_T1\_x10}. The two close
neighbours are masked instead of subtracted (see text for details).
}
\label{fig_x10_msk}
\end{figure*}

\subsubsection{500R\_COSMOS\_T1\_x10}

    This is one of the sources whose Hi$z$Idx
have been revised because of the new ALMA identifications, and yet
still survives the criterion of Hi$z$Idx(500) $\geq 0.7$. The
optical-to-NIR images are shown in Figure \ref{fig_500R_x10}. At the
exact ALMA location, there is no detection in either $I_{814}$ or $H_{160}$.

    The 500~$\mu$m riser counterpart is invisible in all the HSC images. It is 
visible in the UltraVISTA $JHK_s$ images, especially in $K_s$ and $H$. However,
it is not included in the UltraVISTA DR4 catalog. 
We were able to extract this source, together with its two 
neighbours. While these neighbours are not as close
to the target as in the cases of \texttt{500R\_x08} and \texttt{x31},
subtracting them would still benefit the photometry. However, the same GALFIT
procedure used previously was not successful here: the northeastern neighbour
is much brighter than the target, and its subtraction would corrupt the image
at the target location. Therefore, we opted to mask the main bodies of these
two neighbours and carried out the photometry in the masked images. Similar to
the photometry for \texttt{500R\_x08} and \texttt{x31}, we ran SExtractor in 
the dual-image mode using the neighbour-masked $K_s$ image as the detection 
image. The results are listed in Table 2. Figure \ref{fig_x10_msk} shows the 
comparison of the original and the masked images.
Interestingly, only part of the $K_s$ counterpart is visible in $H$ and 
$J$. This could be explained by the uneven distribution of dust. This source is
also prominent in the IRAC images, however it is severely blended with at least
one of its close neighbours. We therefore do not attempt to obtain its IRAC 
photometry.

\subsection{Counterparts of the contaminators}

\subsubsection{500R\_COSMOS\_T1\_x02}

\begin{figure*}
\includegraphics[width=14cm]{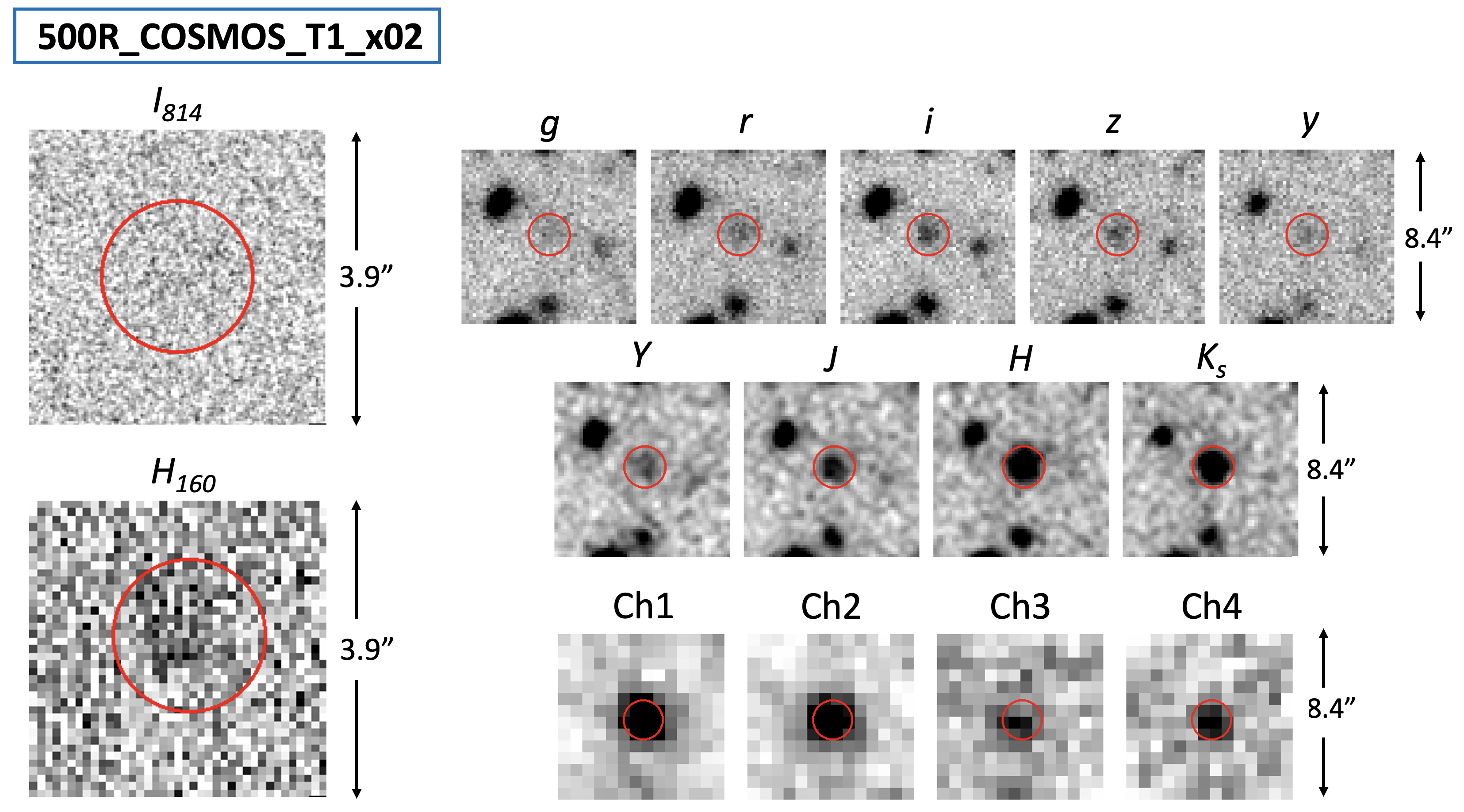}
\caption{Image stamps showing the optical-to-NIR counterpart of
\texttt{500R\_COSMOS\_T1\_x02} and its vicinity. The image arrangement is the
same as in Figure \ref{fig_500R_x08}, and the red circles ($r=1$\arcsec) are 
centred on the ALMA position. It is invisible in $I_{814}$ and is weakly
detected in $H_{160}$. It is clearly detected in all the HSC, VISTA and
IRAC images.
}
\label{fig_500R_x02}
\end{figure*}

   The counterpart identified by the ALMA data
corresponds to a bright VLA source at exactly the same location, which is
6\arcsec.13 away from the HerMES centroid. 
The VLA source was noted in YMHF20, but
it was not taken as the right counterpart due to the large positional 
separation. With the new identification, its Hi$z$Idx(500) is revised to 0.13
and no longer satisfies Hi$z$Idx(500) $>0.7$. Its optical-to-NIR images are
shown in Figure \ref{fig_500R_x02}.

   At the ALMA location, the source is invisible in $I_{814}$ and is very weak 
in $H_{160}$ ($24.02\pm0.29$~mag).
However, it is clearly detected in the HSC $g$ and $r$ bands, which means that
it must be at $z<4$. It is prominent in the other HSC images as well as in the
UltraVISTA images. Its photometry, which
is from the team catalogs, is listed in Table 2. The $K_s$ image is 
significantly more extended than the PSF, and is elongated in the N-S direction.
This suggests that it might have two components.
Moreover, it is detected in IRAC.
In MIPS, its has $S_{24\mu m}$$=$167$\pm$16~$\mu$Jy.

\subsubsection{500R\_COSMOS\_T1\_x34}

    As revealed by the ALMA data, this
source has two counterparts that are separated by 14\arcsec.0, both of which
have VLA counterparts included in the 3~GHz catalog. We designate the one to 
the South-West as the ``34a'' component and the one to the North-East as the
``34b'' component, respectively. The HerMES centroid fall in between these two 
components, and the separation between this centroid to either is larger than 
the criterion adopted in YMHF20. As the result, YMHF20 treated it as with no 
counterpart in 3~GHz, although component 34a was noted in YMHF20 as being a 
close neighbour at 6\arcsec.0 from the HerMES centroid. With the new 
identification, we combine the radio flux densities of both 34a and 34b and 
revise the Hi$z$Idx value of this source to 0.12. 

\begin{figure*}
\includegraphics[width=14cm]{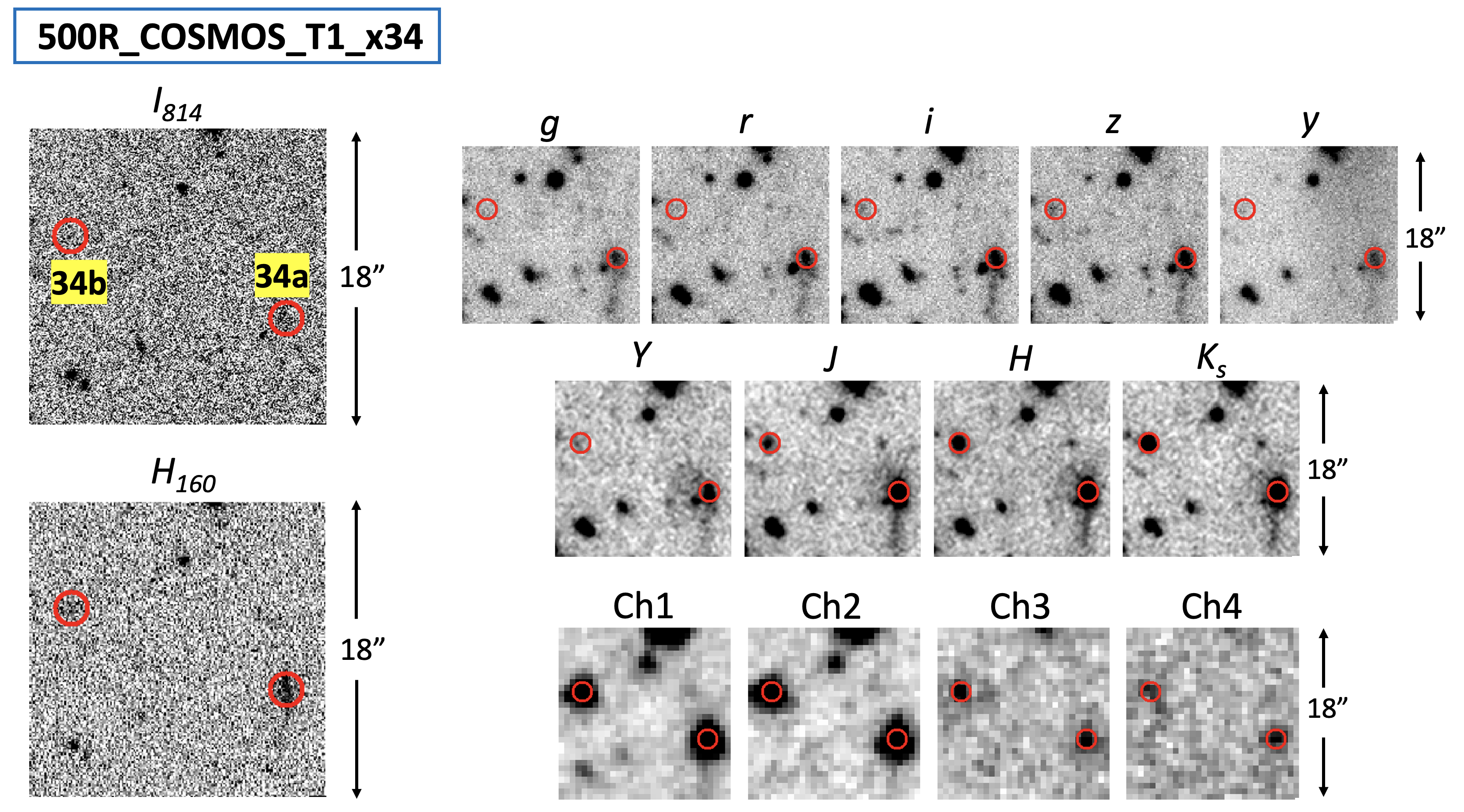}
\caption{Image stamps showing the optical-to-NIR counterpart of
\texttt{500R\_COSMOS\_T1\_x02} and its vicinity. The image arrangement is the
same as in Figure \ref{fig_500R_x08}, but the image sizes are different. This 
source has two components as revealed by ALMA (labeled as ``34a'' and ``34b''),
and the red circles ($r=1$\arcsec) are centred on the ALMA positions. Based on
these images, neither component could be at high-$z$. 
}
\label{fig_500R_x34}
\end{figure*}

   The optical-to-NIR images of this source is shown in 
Figure \ref{fig_500R_x34}.
Component 34a is visible in both $I_{814}$ and $H_{160}$, and our photometry
gives $I_{814}$$=$26.67$\pm$0.22 and $H_{160}$$=$21.60$\pm$0.09~mag, 
respectively.  Component 34b is only visible in $H_{160}$, which we obtained 
$H_{160}$$=$23.35$\pm$0.33~mag. Both components are clearly detected 
in the HSC, the UltraVISTA and the S-COSMOS images.
The quoted photometry in Table 2 are from the respective catalogs in the data
releases.


\subsubsection{SD850\_COSMOS\_T1\_A03}


    At $\sim$20\arcsec\ away from this SCUBA2
source, there is a prominent HerMES source in all three SPIRE bands. This
HerMES source is different from this SPIRE dropout, and in fact has a
different 3~GHz counterpart. For this SPIRE dropout, the ALMA data has 
identified its
counterpart at 3\arcsec.0 from the SCUBA2 centroid. There is a prominent 3~GHz
source at this location, which was noted in YMHF20 but was discarded because
the separation is larger than the adopted criterion. With the new
identification, this 3~GHz source is in fact the radio counterpart; the revised
Hi$z$Idx is reduced to 0.22 and no longer satisfies the high-$z$ criterion of
Hi$z$Idx(850) $>0.5$.

\begin{figure*}
\includegraphics[width=14cm]{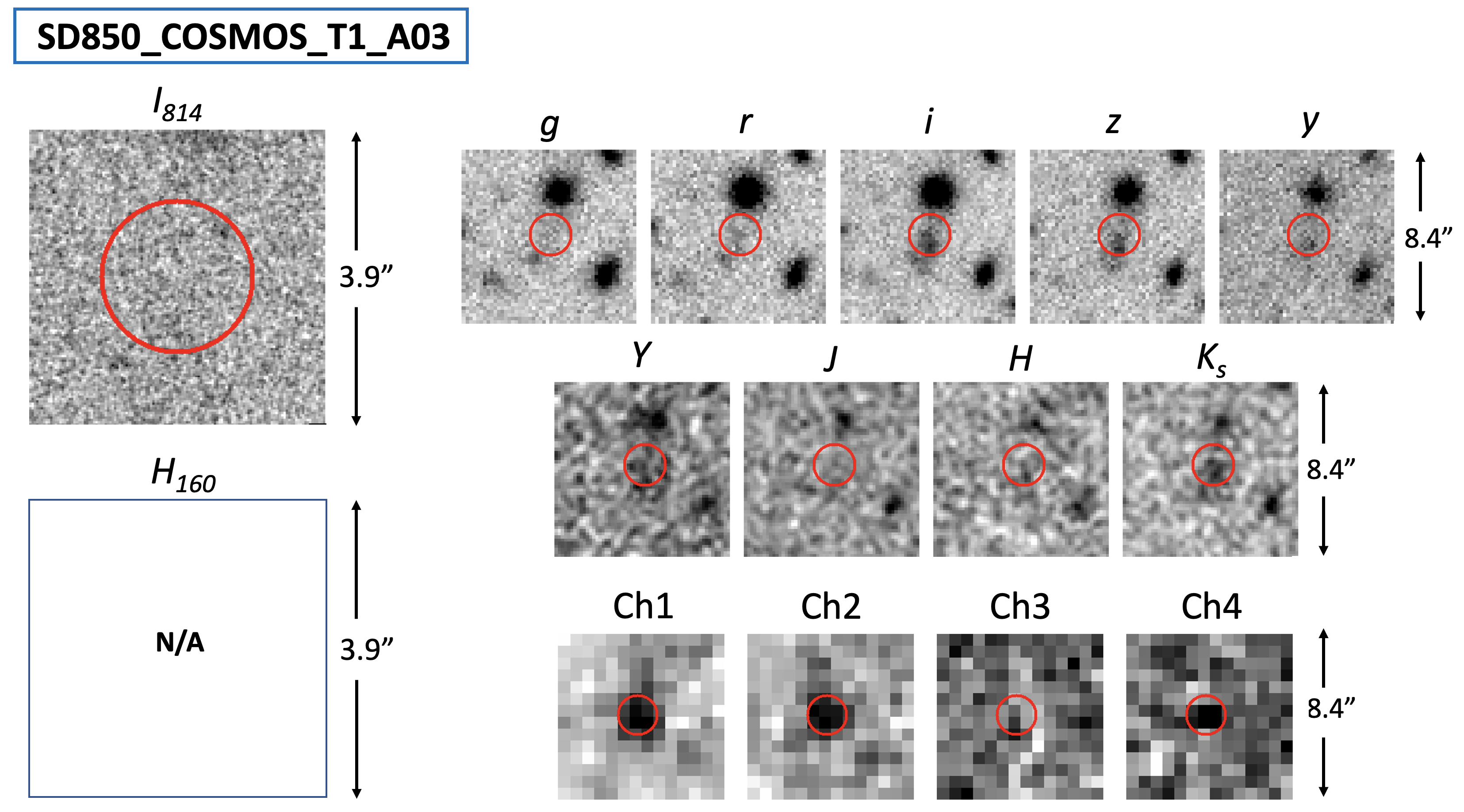}
\caption{Image stamps showing the optical-to-NIR counterpart of
\texttt{SD850\_COSMOS\_T1\_A03} and its vicinity. The image arrangement is the
same as in Figure \ref{fig_500R_x08} except that no $H_{160}$ image is
available. The red circles ($r=1$\arcsec) are centred on the ALMA position.
At this exact location, a faint counterpart is visible in the HSC $riz$ images,
but it is blended with a close, brighter (albeit still faint) neighbour. The 
counterpart is detected in $K_s$ but seems to be blended with a different
close neighbour. There are prominent detections in IRAC Ch1 and Ch2, however it
is unclear how the blended neighbours contribute.
}
\label{fig_SD_A03}
\end{figure*}

   Its optical-to-NIR images are shown in Figure \ref{fig_SD_A03}. The source
is invisible in $I_{814}$. Unfortunately, it does not have $H_{160}$ coverage.
In the HSC images, the source seems to have two very close components that 
cannot be separated for photometry. However, one can still tell that the ALMA
position is closer to the northern component, which is also the fainter one 
among the two in the HSC images. Both components are visible in $r$, and this 
rules out the possibility that this SPIRE dropout could be at $z>5$. In the 
UltraVISTA images, it is only visible in $K_s$, and one can also identify
the two close components as suggested by the HSC images. It is impossible to 
separate these two objects for photometry in any of the current images, and we
treat them as being physically associated.
The quoted photometry in HSC images is taken from the HSC catalog, and include
both components as a single clump. The UltraVISTA catalog does not include
this source; by our photometry, the clump of two components has
$K_s$$=$23.77$\pm$0.16~mag. It is detected in the IRAC Ch1/2 but is invisible in
Ch3/4, and is also invisible in MIPS 24~$\mu$m.
    
\section{Support of the High-z Interpretation}

   Using the multi-band photometry as described above, we constructed the 
optical-to-NIR SEDs for these counterparts. The photometry is listed in 
Table 2, but $I_{814}$, $H_{160}$, Ch3, and Ch4 were not used due to the 
concern that they might introduce large systematic errors in the SED analysis
because of their very different spatial resolutions as compared to the others. 
The analysis included
the objects in Table 1 except two of them: 
(1) \texttt{500R\_COSMOS\_T1\_x26} (``CRLE'' at $z=5.667$) was excluded because
its real counterpart is completely blocked by the foreground galaxy (see 
\S 3.2.3), and (2) \texttt{500R\_COSMOS\_T1\_x24} could not be
included because it is not detected in any of these optical-to-NIR bands.
For \texttt{500R\_COSMOS\_T1\_x31}, we considered two 
scenarios, one with the IRAC photometry and the other
without. We took this approach because it is uncertain whether the IRAC
detections are significantly contaminated by the two neighbours visible in
the UltraVISTA images, a reason detailed in \S 3.2.4.

   We fit these SEDs to the population synthesis models of 
\citet[][``BC03'']{Bruzual2003}. This resulted in their photometric redshifts 
($z_{\rm ph}$) together with some other critical parameters that describe the
underlying stellar populations, such as their stellar masses ($M_*$), their
ages ($T$), etc. All this enables us to further examine whether the 500~$\mu$m
riser hosts could indeed be at high-$z$ based on their Hi$z$Idx values. 

\begin{figure*}
\includegraphics[width=14cm]{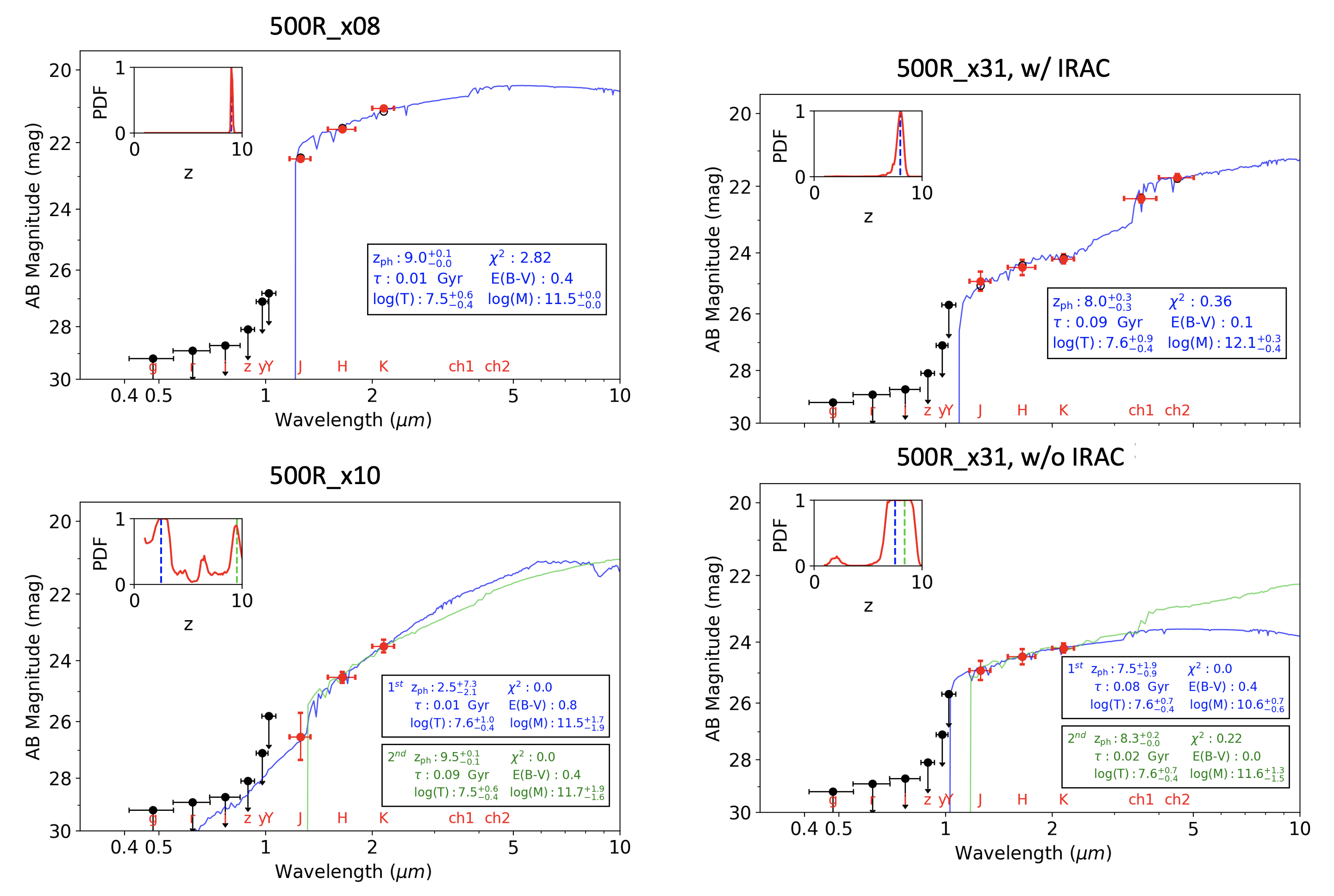}
\caption{Optical-to-NIR SED fitting results for three high-$z$ candidates.
The short ID of each object is indicated in each panel. 
For \texttt{500R\_COSMOS\_T1\_x31}, two scenarios are shown
(one with the IRAC photometry and the other without).
The magnitudes are presented
as red circles, while the limits are shown as black circles with downward
arrows. The relevant passbands are labeled for ease of reading. The inset shows
the PDF of $z_{\rm ph}$, where the dotted blue line indicates the peak that 
corresponds to the best-fit $z_{\rm ph}$. When multiple peaks exist, the
second peak is indicated in dotted green line. The blue curve is the best-fit
model spectrum. The best-fit model corresponds to the secondary $z_{\rm ph}$, 
if exists, is shown in green. The fitted parameters are labeled in the boxes,
where those associated with the first and the second $z_{\rm ph}$ peaks are
coded in blue and green, respectively. The age ($T$, in yr) and the stellar 
mass ($M_*$, in $M_\odot$) are in logarithmic scale. 
}
\label{fig_oirsed_highz}
\end{figure*}

\begin{figure*}
\includegraphics[width=14cm]{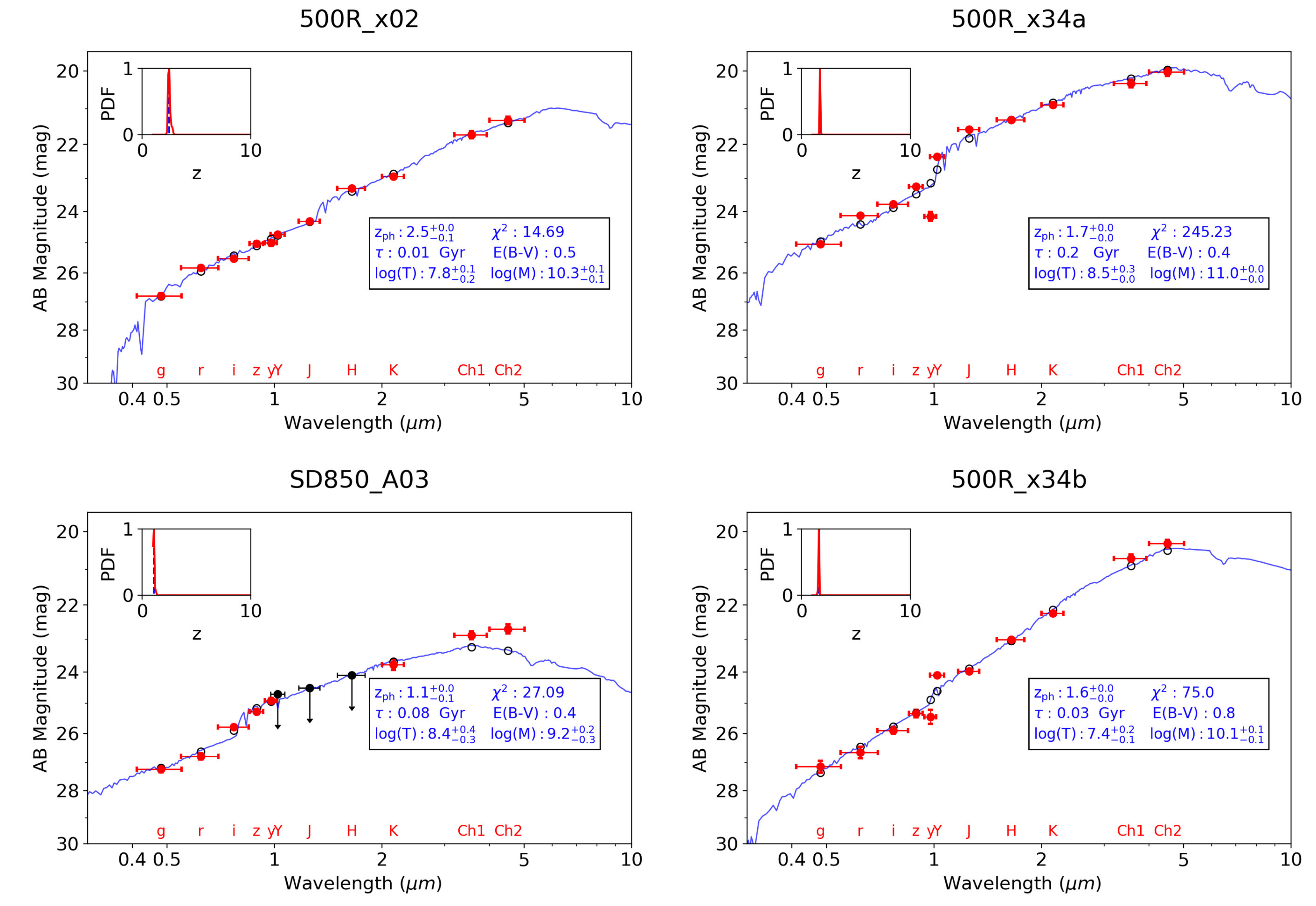}
\caption{Similar to Figure \ref{fig_oirsed_highz}, but for the contaminators.
The small open circles are the simulated magnitudes (based on the best-fit
models) in the corresponding bands.
Object \texttt{500R\_COSMOS\_T1\_x34} has two components (``a'' and ``b''),
and they are fitted separately. The large $\chi^2$ in both cases are driven
by the descrepancy in the photometry of the HSC-SSP $y$-band and the UltraVISTA
$Y$-band.
}
\label{fig_oirsed_contaminator}
\end{figure*}

\subsection{Host galaxy SED fitting process}

   We used {\it Le Phare}\ \citep[]{Arnouts1999, Ilbert2006} to fit the SEDs.
The system response curves in various bands were taken from the public websites
of the corresponding instruments. We adopted a series of BC03 composite stellar
population models constructed using simple stellar populations of solar 
metalliticity and the initial mass function (IMF) of \citet[][]{Chabrier2003}.
These composite models follow exponentially declining star formation histories,
with the characteristic time scale $\tau$ varying from 10~Myr to 13~Gyr. We 
further assumed the dust extinction law of \citet[][]{Calzetti1994} and allowed
$E(B-V)$ to vary from 0 to 1.5. The range of redshift was from $z=0$ to $z=10$.
{\it Le Phare}\ uses the formalism of \citet[][]{Madau1995} to account for the 
absorption due to intervening neutral hydrogen (H I) clouds, and also limits 
the age of the fit model to be smaller than the age of the universe at the 
fitted redshift. The step-size
in redshift was set to $\Delta z=0.1$. {\it Le Phare} also allows adjustments
of photometric errors to account for the possible systematic errors across 
different instruments. For this purpose, we added in quadrature 0.1~mag
to the photometric errors of the IRAC bands and 0.05~mag to the others. To
guard against obviously inappropriate fits, we also applied a loose prior
that the absolute magnitude in $K_s$ must be within $-10$ and $-26$~mag. 

   {\it Le Phare}\ takes the minimum $\chi^2$ approach in doing fitting, and 
produces a probability distribution function over the allowed redshift range 
(PDF; $\propto e^{-0.5\chi^2_{min}(z)}$). The redshift at which the PDF 
has its peak (i.e., the lowest $\chi^2_{min}(z)$) is deemed as the best-fit 
$z_{\rm ph}$, and its errors are computed using the redshifts where 
$\chi^2_{min}$ are $\pm 1$ from the lowest value. The model that results in the
lowest $\chi^2_{min}$ at this redshift is what {\it Le Phare}\ calls as the
best-fit model, and its associated parameters (such as $M_*$, $T$, $\tau$ etc.)
are the best-fit values of these parameters. The common practice, however, is
to adopt the median values instead of the best-fit values for these other
parameters, because {\it Le Phare}\ generates errors associated with the median
but not the best-fit values. We also adopted this common practice. A slight
complication would occur when the PDF has more than one peak. In this case,
we ran the software for the second time, which was to fix the redshift to the
secondary peak to derive other parameters around this secondary solution.

    The results are summarized in Figure \ref{fig_oirsed_highz} and 
\ref{fig_oirsed_contaminator} for the high-$z$ candidates
(\texttt{500R\_COSMOS\_T1\_x08}, \texttt{x31} and \texttt{x10}) and the
contaminators (\texttt{500R\_COSMOS\_T1\_x02}, \texttt{x34} and 
\texttt{SD850\_COSMOS\_T1\_A03}), respectively. Obviously, the analysis for 
the high-$z$ sample suffers from the limitted number of passbands that have
positive detections. However, the non-detection bands are deep enough that they
play a critcal role in narrowing down the possible solutions. While the
$z_{ph}$ estimates cannot be taken as the confirmations, they do lend support
that the Hi$z$Idx(500) criterion works:
\texttt{500R\_COSMOS\_T1\_x08} has $z_{\rm ph}=9.0$; \texttt{x31} has
$z_{\rm ph}=8.0$ and 7.5 in the scenarios with and without the IRAC
photometry, respectively; \texttt{x10} has a rather chaotic PDF whose first 
peak is at $z_{\rm ph}=2.5$, but it still has
the second peak at $z_{\rm ph}=9.5$.
In other words, the objects in the
high-$z$ sample all have acceptable solutions at $z>6$.

   In contrast, the contaminators all have best-fit $z_{\rm ph}$ at 
$z\approx 1$--3 where the bulk of FIR/sub-mm sources lie. 
For \texttt{500R\_COSMOS\_T1\_x34}, its two components (``a'' and ``b'') were
fitted independently and the results are also shown as such; it is very likely
that they are at the same redshift and therefore physically associated,
however the conclusion is not yet definite.

\begin{figure*}
\includegraphics[width=14cm]{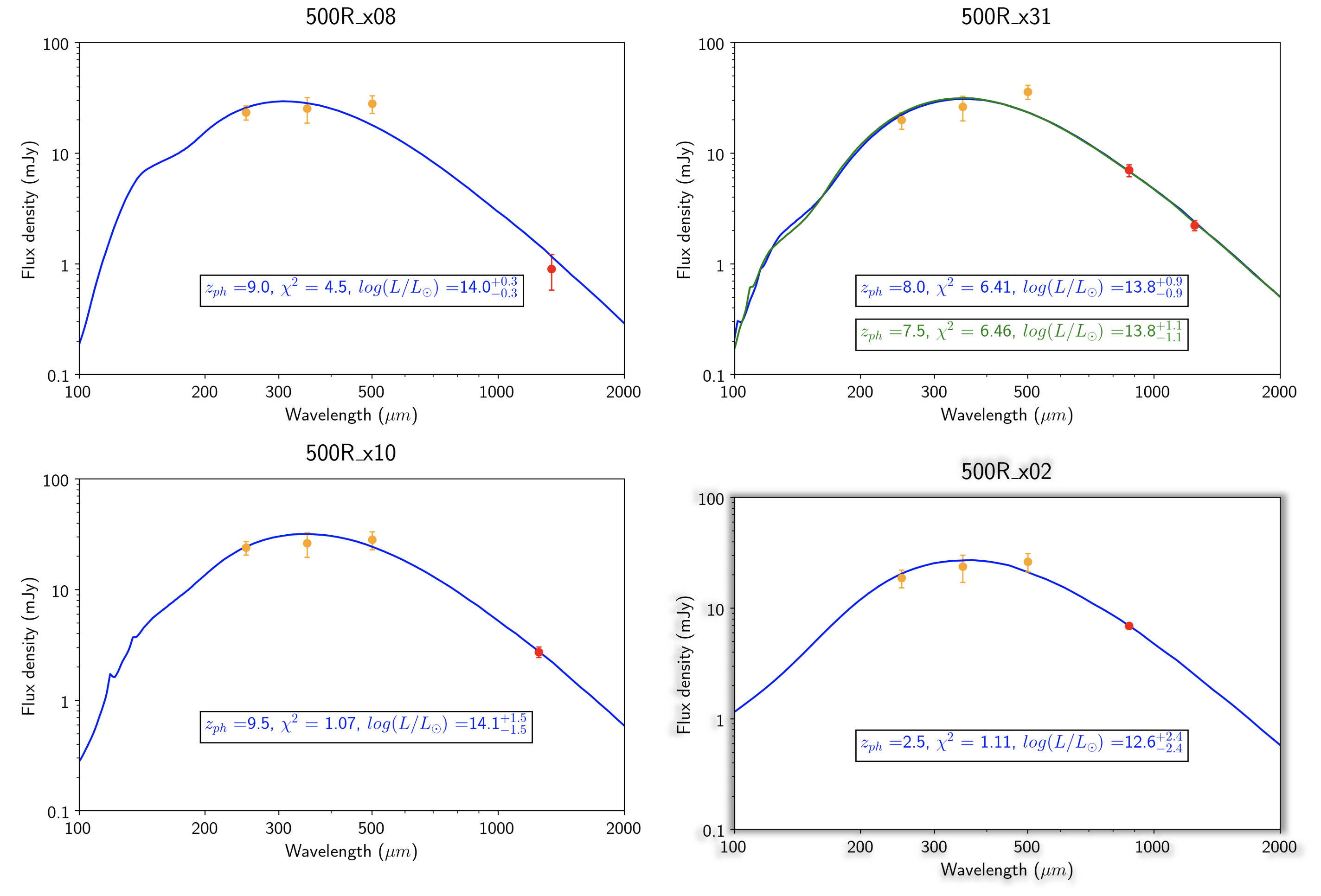}
\caption{FIR-to-mm SED fitting results for three high-$z$ candidates (plots
without shadow) and one contaminator (plot with shadow). The SPIRE and ALMA 
photometry are shown in orange and red symbols, respectively, which are fitted
to the models of \citet[][]{Siebenmorgen2007} at the fixed $z_{\rm ph}$ values
as described in \S 4.1. The best-fit model spectra are superposed, and the
derived $L_{\rm IR}$ values are labeled. 
For \texttt{500R\_COSMOS\_T1\_x10},
only the fit at $z_{\rm ph}=9.5$ is shown. For \texttt{500R\_COSMOS\_T1\_x31},
the fits correspond to the first $z_{\rm ph}$ peaks in both scenarios (with
and without IRAC photometry when deriving $z_{\rm ph}$) are shown (in blue and 
green, respectively). 
}
\label{fig_firsed}
\end{figure*}

\subsection{Total IR luminosity and instantaneous SFR}

   Using the derived $z_{\rm ph}$, we fit the FIR-to-mm SEDs of our objects
to the starburst models of \citet[][]{Siebenmorgen2007} to obtain their total
IR luminosity ($L_{\rm IR}$; integrated over rest-frame 8--1000~$\mu$m),
following \citet[][]{MY2015}. The SPIRE photometry is taken from YMHF20.
The ALMA photometry (see Table 1) are incorporated. This analysis includes 
three objects in the high-$z$ sample: \texttt{500R\_COSMOS\_T1\_x08}, 
\texttt{x31} (both of its $z_{\rm ph}$ peak solutions are used) and \texttt{x10}
(its secondary $z_{\rm ph}$ peak solution is used). It also include one object
(\texttt{500R\_COSMOS\_T1\_x02}) in the
contaminator sample. We do not fit the other two contaminators: 
\texttt{500R\_COSMOS\_T1\_x34} is made of two components and we cannot obtain
their fluxes separately in their SPIRE bands, and 
\texttt{SD850\_COSMOS\_T1\_A03} only has photometry at the Rayleigh-Jeans tail.

   Figure \ref{fig_firsed} shows the fitting results. 
For \texttt{500R\_COSMOS\_T1\_x31}, its $z_{\rm ph}$ peak solutions in both 
scenarios are used. For \texttt{500R\_COSMOS\_T1\_x10}, its secondary
$z_{\rm ph}$ peak solution is used. Without exception, all the three high-$z$
candidates have $L_{\rm IR} >10^{13} L_\odot$: 
\texttt{500R\_COSMOS\_T1\_x08}, \texttt{x31} and \texttt{x10} have 
$L_{\rm IR}=1.0\times 10^{14}$, 6.3 $\times 10^{13}$ and 
$1.3\times 10^{14} L_\odot$, respectively. Applying the standard conversion
given by \citet[][]{Kennicutt1998b}, which is
${\rm SFR_{IR}}=1.0\times 10^{-10}L_{\rm IR}$ after adjusting for a Chabrier 
IMF, the inferred instantaneous SFRs are $10^4$, 6.3 $\times 10^3$ and
$1.3\times 10^4 M_\odot$~yr$^{-1}$, respectively.

\begin{figure*}
\includegraphics[width=18cm]{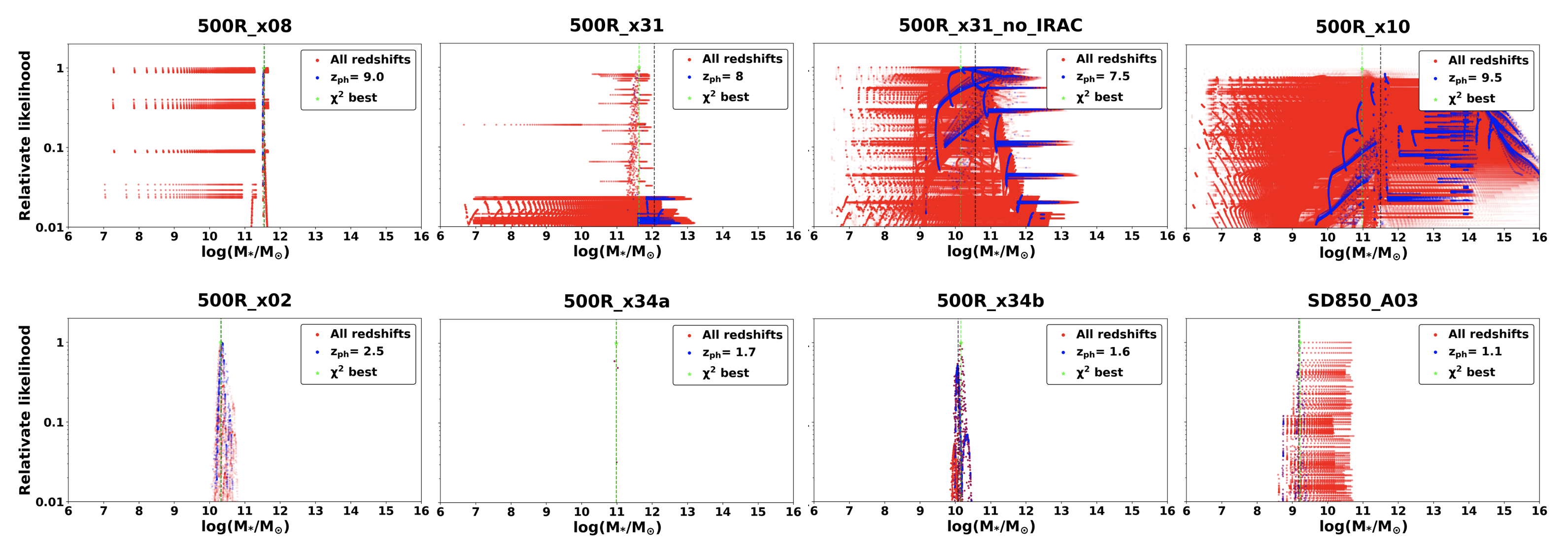}
\caption{Relative likelihood distributions of derived stellar masses. The top
three panels are for the high-$z$ candidates in Figure \ref{fig_oirsed_highz}
and the bottom four panels are for the contaminators in Figure
\ref{fig_oirsed_contaminator}. The red dots are for the fitted templates at
all redshifts, while the blue dots are for the fits at the labeled $z_{\rm ph}$.
The large filled, green (yellow) circles represent the corresponding values of
the first (second) $z_{\rm ph}$ peaks. The fits of the contaminators all have
reasonably well defined, approximately log-normal distributions at the adopted 
$z_{\rm ph}$, thanks to the large number of passbands with positive detections.
In contrast, the fits of the high-$z$ objects have very chaotic distributions
due to the limited passbands with positive detections, which makes their
derived stellar masses less trustworthy.
}
\label{fig_chi2mass}
\end{figure*}

\subsection{Stellar masses: potential problem or not}

    While the counterpart SED fitting gives $z_{\rm ph}>6$ for the high-$z$
candidates, there could be a potential problem with the inferred stellar
populations. The high-$z$ solutions of \texttt{500R\_COSMOS\_T1\_x08}, 
\texttt{x31} and \texttt{x10} report stellar masses of $10^{11.5}$, $10^{12.1}$
($10^{10.6}$) and $10^{11.7} M_\odot$ at $z_{\rm ph}=9.0$, 8.0 (7.5) and
9.5, respectively. These are extremely high values, previously unseen in such
early epochs (and in fact very rare in any epochs). At these redshifts,
the age of the universe was only $\sim$555, 651 (709) and 516~Myr, 
respectively. If the epoch of first stars was at $z=17.2$ 
\citep[][]{Bowman2018}, i.e., $\sim$225~Myr after the Big Bang, there would
only be $\sim$330, 426 (484) and 291~Myr at most, respectively, for them to 
assemble such huge stellar masses, and therefore their average SFRs over these
periods would have to be $\sim$958, 2,955 (82) and 1,722~$M_\odot$~yr$^{-1}$,
respectively. To make the situation
worse, the best-fit models all have very small $\tau$ values of 10--90~Myr,
which means that {\it almost all their stars}\ are formed instantaneously at
the observed redshifts, a scenario that is to the opposite of the hierarchical
formation picture of high-mass objects.

    However, we should point out that these inferred parameters need to be
treated with caution. As explained in \S 4.1, the reported $log(M_*)$ (and 
other parameters such as $\tau$ as well) value is the median value of all the
fits weighted by the relative likelihood ($\propto 1/\sqrt{e^{\chi^2}}$). This
would work if the relative likelihood function behaves reasonably well. 
Unfortunately, it is not the case for these high-$z$ objects. Figure 
\ref{fig_chi2mass} shows the relative likelihood distributions of $M_*$ for
both the high-$z$ objects and the contaminators. It is obvious that the
distributions for the contaminators are approximately log-normal while those
for the high-$z$ objects are completely chaotic. The major cause of the
difference is that the contaminators all have at least eight bands of 
detections while the high-$z$ objects only have three to five bands. This has
made the fits of the high-$z$ objects not well constrained. In these cases,
the ``median'' values lose the physical meaning. 

    Nevertheless, it is also worth pointing out that the best-fit models still
give rather high stellar masses for the high-$z$ objects. This is not 
unexpected from their optical-to-near-IR photometry, because the bright observed
magnitudes in the near-IR indeed would imply such high stellar masses under any
reasonable assumption of the mass-to-light ratio. In other words, if we accept
the high $z_{\rm ph}$ for these objects, we will have to accept that the
aforementioned tension still exists (albeit to a lesser degree) between the 
huge stellar mass and the limited time to assemble stars.

   One might argue that these sources could have significant AGN contribution
and therefore fitting the SEDs to stellar population synthesis models is not
appropriate. However, there is no indication that these sources host AGN. The
COSMOS field has deep X-ray observations from the Chandra Cosmos Legacy Survey
\citep[][]{Civano2016}, and none of our sources are detected in these data.

   There is an alternative that would make the situation less problematic. If
these objects are gravitationally lensed by a large factor (e.g., 
$\mu \sim 10$), the intrinsic stellar masses would then be smaller by the
same factor and therefore the tension will be reduced
(although still not eliminated). All the three high-$z$
objects have very close foreground companions (see Figure \ref{fig_500R_x08},
\ref{fig_500R_x31} and \ref{fig_500R_x10}), making
gravitational lensing an attractive explanation. However, we will not be able
to have a definite answer until spectroscopic redshifts are obtained for the
counterparts and the companions.

\section{Discussion}

    Despite the difficulty with $M_*$ as mentioned above, the derived 
$z_{\rm ph}$ for the high-$z$ objects should have better reliability because 
these largely depend on the locations of the Lyman-break signature, which is 
determined by the intervening H I absorption but not the intrinsic properties
of the galaxies.  Trusting their high-$z$ solutions in \S 4.1,
we can estimate the surface density of ULIRGs at $z>6$. 
In YMHF20, there are a total of 17 Tier 1 500~$\mu$m risers in the 
COSMOS field that fall within the coverage (2~deg$^2$) of the VLA 3~GHz data.
Based on the revised matching criterion as described in \S 3.1, nine of them
have Hi$z$Idx(500) $\geq 0.7$. We are able to study five of these nine in this
work, thanks to the A$^3$COSMOS positions. Among these five, only one 
(\texttt{500R\_COSMOS\_T1\_x26}; ``CRLE'' at $z=5.667$) is 
definitely at $z<6$, three (\texttt{500R\_COSMOS\_T1\_x08}, \texttt{x10} and 
\texttt{x31}) have optical-to-NIR counterparts consistent with being at $z>6$,
and one (\texttt{500R\_COSMOS\_T1\_x24}) is inconclusive due to the lack of 
optical-to-NIR counterpart. Therefore, the success rate of 
Hi$z$Idx(500) $\geq 0.7$
selection among 500~$\mu$m risers is 60--80\% (three or four out of five,
depending on whether \texttt{500R\_COSMOS\_T1\_x24} is counted as a legitimate
high-$z$ candidate).
The overall success rate of YMHF20's 500~$\mu$m risers in selecting $z>6$ 
objects is 32--42\% (nine out of seventeen and multiplied by 60\% or 80\%).
The surface density of $z>6$ dusty starbursts selected as 500~$\mu$m risers is 
therefore 2.7--3.6~deg$^{-2}$.
On the other hand, the SPIRE dropouts so far do not seem to be able to select 
any objects at $z>6$, as no such object satisfies Hi$z$Idx(850) $\geq 0.5$ 
(\texttt{SD850\_COSMOS\_T1\_A03} has the revised Hi$z$Idx(850) $=0.22$).

    As already discussed, accepting the three objects shown in Figure 
\ref{fig_oirsed_highz} being at $z>6$ also means that we probably need to 
accept the high stellar masses of their host galaxies, which are at
$\gtrsim 10^{11} M_\odot$ if they are not gravitationally lensed. 
The possible formation scenarios have two extremes: they
kept forming stars at an average rate of hundreds or even a few thousand
$M_\odot$~yr$^{-1}$, or they formed all their stars instantaneously through 
enormous bursts. While such very high-mass, high-SFR objects at high-redshifts 
are at odd with the currently accepted picture of the early star formation, 
there are now theoretical arguments that their existence is needed to explain
the supermassive black holes in the early universe \citep[][]{Kroupa2020}.
Regardless of their detailed formation histories, the progenitors of such
objects must once be very luminous in the rest-frame UV, i.e., before
they were heavily enriched by metal and formed large amount of dust. For
example, using the conversion between $L_{\rm UV}$ and SFR as formulated in 
\citet[][]{Madau1998}, SFR $=500$~$M_\odot$~yr$^{-1}$ would translate to
$M_{\rm UV}=-24.9$~mag. Such a progenitor, if observed at $z=12$, would show
up as a bright $H$-band dropout with $K \sim 22.9$~mag.
\footnote{We note that the same argument 
was presented in \citet[][]{Yan2006} for the progenitors of high stellar 
mass galaxies observed at $z\approx 6$, although the implication there was less
extreme because those $z\approx 6$ galaxies have stellar masses on the order of
several $10^{10} M_\odot$.}. 
If we accept the surface density derived above, we should expect a similar
surface density of their progenitors that manifest themselves as
bright $K$-band objects.

\section{Summary}

    In this work, we study a subsample of the high-$z$ ULIRG candidates
selected by YMHF20 as 500~$\mu$m risers and SPIRE dropouts. Our objects are in 
the COSMOS field, where the archival ALMA data from the A$^3$COSMOS program
offer the opportunity to pinpoint their exactly locations. We aim at the 
candidates at $z>6$, which are those so weak in radio that they meet the 
Hi$z$Idx criteria for $z>6$. In total, our sample includes seven 500~$\mu$m
risers and one SPIRE dropout. Based on the ALMA positions, we have found that
the matching criteria adopted by YMHF20 between the FIR/sub-mm and the radio 
positions are too stringent and need to be relaxed slightly. Based on the 
subsequent revision of their Hi$z$Idx values, the eight objects form the true 
high-$z$ subsample at $z>6$ consisting of five objects and the contaminator 
subsample consisting of three objects (including the SPIRE dropout).
We searched for their counterparts in the deep optical-to-NIR images
available in the field, and carried out SED analysis of the host galaxies using
population synthesis model. The objects in the contaminator sample all have
optical-to-NIR SEDs consistent with being interloppers at $z\sim 1$--3. In the
high-$z$ subsample, one object turns out to be a known galaxy at $z=5.667$, one
has no visible counterpart in any of the optical-to-NIR images, and the other
three have solutions with $z_{\rm ph} > 6$. If we trust this assessment, 
the overall success rate of YMHF20's 500~$\mu$m riser in selecting
dusty starbursts at $z>6$ is $\sim$32--42\%, and the success rate increases to
$\sim$60--80\% after the purification by applying the Hi$z$Idx criterion. The
surface density of $z>6$ 500~$\mu$m risers is $\sim$2.7--3.6~deg$^{-2}$. The
very existence of dusty starburst at $z>6$ implies that the universe must have
been actively forming stars very early in time so that dust could be present at
the redshifts where these ULIRGs are observed. The inferred stellar masses of
their host galaxies also suggest that their progenitors could have been in
starburst state since formation. Spectroscopic confirmation of such objects,
both in the millimeter regime and in the near-IR regime, will be critical in
understanding the star formation processes in the very early universe.

\section*{Acknowledgements}

  HY and CL acknowledge the support of the University of Missouri Research 
Council Grant URC-21-005. We thank all the teams that collected extensive data
in the COSMOS field and put the data to the public domain. 
The accurate positioning is made possible by the A$^3$COSMOS program. 
We utilize the UltraVISTA data, which are based on observations
made with ESO Telescopes at the La Silla or Paranal Observatories under
program ID(s) 179.A-2005(A), 179.A-2005(B), 179.A-2005(C), 179.A-2005(D),
179.A-2005(E), 179.A-2005(F), 179.A-2005(G), 179.A-2005(H), 179.A-2005(I),
179.A-2005(J), 179.A-2005(K). We also make use of data collected at the Subaru
Telescope and retrieved from the HSC data archive system, which is operated by
the Subaru Telescope and Astronomy Data Center (ADC) at NAOJ.

\section*{Data Availability}

The data underlying this article are available in the article.



\bibliographystyle{mnras}






\bsp    
\label{lastpage}
\end{document}